%% file: sn-article.tex
\PassOptionsToPackage{table}{xcolor}
\documentclass[default]{sn-jnl}% Default
\usepackage{graphicx}%
\usepackage{multirow}%
\usepackage{amsmath,amssymb,amsfonts}%
\usepackage{amsthm}%
\usepackage{mathrsfs}%
\usepackage[title]{appendix}%
\usepackage{xcolor}%
\usepackage{textcomp}%
\usepackage{manyfoot}%
\usepackage{booktabs}%
\usepackage{algorithm}%
\usepackage{algorithmicx}%
\usepackage{algpseudocode}%
\usepackage{listings}%
\usepackage{caption}

\usepackage[table]{xcolor}
\definecolor{cusyellow}{HTML}{D8D6C2}
\definecolor{cusyellowl}{HTML}{EBEADE}

\begin{document}

\title[Generalizable Cervical Cancer Screening via Large-scale Pretraining and Test-Time Adaptation]{\fontsize{14}{16}\selectfont Generalizable Cervical Cancer Screening via Large-scale Pretraining and Test-Time Adaptation}

\author[1]{\fnm{Hao} \sur{Jiang}}\email{hjiangaz@connect.ust.hk}\equalcont{These authors contributed equally to this work.}

\author[1]{\fnm{Cheng} \sur{Jin}}\email{cjinag@connect.ust.hk}\equalcont{These authors contributed equally to this work.}

\author[2]{\fnm{Huangjing} \sur{Lin}}\email{hjlin@cse.cuhk.edu.hk}\equalcont{These authors contributed equally to this work.}

\author[2,3]{\fnm{Yanning} \sur{Zhou}}\email{amandayzhou@tencent.com}

\author[2]{\fnm{Xi} \sur{Wang}}\email{xiwang@cse.cuhk.edu.hk
}

\author[1]{\fnm{Jiabo} \sur{Ma}}\email{jmabq@connect.ust.hk}

\author[4]{\fnm{Li} \sur{Ding}}\email{dingli6@mail.sysu.edu.cn}

\author[5,6,7]{\fnm{Jun} \sur{Hou}}\email{houjun0709@126.com}

\author[1]{\fnm{Runsheng} \sur{Liu}}\email{rliuar@connect.ust.hk}

\author[1]{\fnm{Zhizhong} \sur{Chai}}\email{zchaiab@connect.ust.hk}

\author[1,8]{\fnm{Luyang} \sur{Luo}}\email{luyang\_luo@hms.harvard.edu}

\author[4]{\fnm{Huijuan} \sur{Shi}}\email{shihj@mail.sysu.edu.cn}

\author[9]{\fnm{Yinling} \sur{Qian}}\email{yl.qian@siat.ac.cn}

\author[9]{\fnm{Qiong} \sur{Wang}}\email{wangqiong@siat.ac.cn}

\author*[5,6,7]{\fnm{Changzhong} \sur{Li}}\email{lichangzhong@163.com}

\author*[4]{\fnm{Anjia} \sur{Han}}\email{hananjia@mail.sysu.edu.cn}

\author*[10]{\fnm{Ronald Cheong Kin} \sur{Chan}}\email{ronaldckchan@cuhk.edu.hk}

\author*[1,11,12,13,14]{\fnm{Hao} \sur{Chen}}\email{jhc@cse.ust.hk}

\affil[1]{\orgdiv{Department of Computer Science and Engineering}, \orgname{The Hong Kong University of Science and Technology}, \orgaddress{\city{Hong Kong SAR}, \country{China}}}

\affil[2]{\orgdiv{Department of Computer Science and Engineering}, \orgname{The Chinese University of Hong Kong}, \orgaddress{\city{Hong Kong SAR}, \country{China}}}

\affil[3]{\orgname{Tencent AI Lab}, \orgaddress{\state{Shenzhen}, \country{China}}}

\affil[4]{\orgdiv{Department of Pathology}, \orgname{The First Affiliated Hospital, Sun Yat-sen University}, \orgaddress{\state{Guangzhou}, \country{China}}}

\affil[5]{\orgdiv{Center of Obstetrics and Gynecology}, \orgname{Peking University Shenzhen Hospital}, \orgaddress{\state{Shenzhen}, \country{China}}}

\affil[6]{\orgdiv{Institute of Obstetrics and Gynecology}, \orgname{Shenzhen PKU-HKUST Medical Center}, \orgaddress{\state{Shenzhen}, \country{China}}}

\affil[7]{\orgname{Shenzhen Key Laboratory on Technology for Early Diagnosis of Major Gynecologic Diseases}, \orgaddress{\state{Shenzhen}, \country{China}}}

\affil[8]{\orgdiv{Department of Biomedical Informatics}, \orgname{Harvard University}, \orgaddress{\country{USA}}}

\affil[9]{\orgdiv{Shenzhen Institute of Advanced Technology}, \orgname{Chinese Academy of Sciences}, \orgaddress{\state{Shenzhen}, \country{China}}}

\affil[10]{\orgdiv{Department of Anatomical and Cellular Pathology}, \orgname{The Chinese University of Hong Kong}, \orgaddress{\city{Hong Kong SAR}, \country{China}}}

\affil[11]{\orgdiv{Department of Chemical and Biological Engineering}, \orgname{The Hong Kong University of Science and Technology}, \orgaddress{\city{Hong Kong SAR}, \country{China}}}

\affil[12]{\orgname{HKUST Shenzhen-Hong Kong Collaborative Innovation Research Institute}, \orgaddress{\state{Shenzhen}, \country{China}}}

\affil[13]{\orgdiv{Division of Life Science}, \orgname{The Hong Kong University of Science and Technology}, \orgaddress{\city{Hong Kong SAR}, \country{China}}}

\affil[14]{\orgdiv{State Key Laboratory of Molecular Neuroscience}, \orgname{The Hong Kong University of Science and Technology}, \orgaddress{\city{Hong Kong SAR}, \country{China}}}

\clearpage

\abstract{Cervical cancer is a leading malignancy in female reproductive system. While AI-assisted cytology offers a cost-effective and non-invasive screening solution, current systems struggle with generalizability in complex clinical scenarios. To address this issue, we introduced \textbf{Smart-CCS}, a generalizable \textbf{C}ervical \textbf{C}ancer \textbf{S}creening paradigm based on pretraining and adaptation to create robust and generalizable screening systems.
To develop and validate Smart-CCS, we first curated a large-scale, multi-center dataset named CCS-127K, which comprises a total of 127,471 cervical cytology whole-slide images collected from 48 medical centers. 
By leveraging large-scale self-supervised pretraining, our CCS models are equipped with strong generalization capability, potentially generalizing across diverse scenarios. Then, we incorporated test-time adaptation to specifically optimize the trained CCS model for complex clinical settings, which adapts and refines predictions, improving real-world applicability. We conducted large-scale system evaluation among various cohorts. In retrospective cohorts, Smart-CCS achieved an overall area under the curve (AUC) value of 0.965 and sensitivity of 0.913 for cancer screening on 11 internal test datasets. In external testing, system performance maintained high at 0.950 AUC across 6 independent test datasets. In prospective cohorts, our Smart-CCS achieved AUCs of 0.947, 0.924, and 0.986 in three prospective centers, respectively. Moreover, the system demonstrated superior sensitivity in diagnosing cervical cancer, confirming the accuracy of our cancer screening results by using histology findings for validation. Interpretability analysis with cell and slide predictions further indicated that the system's decision-making aligns with clinical practice.
Smart-CCS represents a significant advancement in cervical cancer screening and highlights the potential for generalizable screening in real-word practice across diverse clinical contexts.}

\maketitle

\section*{Introduction}\label{sec1}
\input{sec/sec1.tex}

\section*{Results}\label{sec2}
\input{sec/sec2.tex}

\section*{Discussion}\label{sec3}
\input{sec/sec3.tex}

\section*{Methods}\label{sec4}
\input{sec/sec4.tex}

\input{sec/sec5.tex}

\bibliographystyle{unsrt}
\bibliography{sn-bibliography}

\begin{appendices}
\input{sec/sec6.tex}
\end{appendices}

\end{document}

%% file: sec/sec1.tex
Cervical cancer, which originates in the cervix, is the fourth most common cancer among women \cite{siegel2024cancer, siegel2023cancer, cohen2019cervical, denny2024cervical}. 
To combat this global health challenge, the World Health Organization (WHO) has announced the cervical cancer elimination initiative, with the objective of 70\% screening coverage among women aged 35-45 worldwide \cite{world2020global}.
Besides, the 5-year relative survival rate for early-stage cervical cancer is 91\%, while the rate significantly drops to 19\% for the late metastatic stage \cite{ref_NCI1}. Therefore, cervical cancer is highly preventable and curable, with early detection being crucial for effective treatment. 

Cytological examination provides a non-invasive, effective, and cost-efficient alternative, making it particularly suitable for widespread screening of precancerous conditions. It is recommended for the early detection of precancerous neoplasia and cervical cancer \cite{kim2023smart,ouh2021discrepancy}. The typical workflow of cervical cancer screening (CCS) is shown in Extended Data Fig. \ref{F1_intro}(a), where cytologists examine cervical cells and identify suspicious cells using a microscope or digital slides. Reporting of cervical cytology results is guided by the Bethesda System (TBS) \cite{nayar2015bethesda}, which provides widely accepted guidelines for standardized manual interpretation. 

However, several challenges affect the efficacy of manual CCS, summarized in Extended Data Fig. \ref{F1_intro}(b). First, cell-level identification is inherently challenging due to cytomorphology similarities with different cell types (squamous and glandular), locations (superficial, intermediate, and basal), and neoplasia (metaplastic, koilocytotic, dyskeratotic) \cite{plissiti2018sipakmed}. Second, each specimen typically contains 20,000 to 50,000 cells with sparsely distributed lesion cells \cite{lin2021dual}, and digitized slides can measure up to 100,000 $\times$ 100,000 pixels. Therefore, examining cytology specimens is tedious and time-consuming for cytologists. Third, patient-level screening results can be significantly affected by the specimen quality and cytologist experience, which can lead to lower reproducibility and objectivity.

With recent advancements, artificial intelligence (AI) is expected to revolutionize the field of computational cytology \cite{jiang2023deep}. Leveraging advanced neural network architectures such as CNNs, RNNs, and Transformers, AI has achieved promising results in several typical cytology tasks, including classification (e.g., HErlev \cite{jantzen2005pap}, SIPaKMeD \cite{plissiti2018sipakmed}), detection (e.g., CERVIX93 \cite{phoulady2018new}), and segmentation (e.g., ISBI \cite{lu2015improved}, CPS \cite{jiang2023donet,liu2024gains}). Even beyond these tasks, recent studies have demonstrated the efficacy of AI-powered cytology analysis in tumor identification and origin prediction \cite{tian2024prediction}.

In the context of AI-assisted cervical cancer screening (CCS), whole slide image (WSI) analysis has emerged as a promising solution through abnormal cell identification and slide-level classification. Unlike histology, where tissues are continuous and regional, cytology involves discrete cell objects and scattered diagnostic clues, necessitating tailored WSI analysis approaches \cite{jiang2023systematic}. Early efforts, such as the dual-path network for cell-level prediction and rule-based WSI classifiers \cite{lin2021dual}, laid the foundation in this field. 
Following this, subsequent studies incorporated more specialized knowledge to provide enriched cytology characteristics and guidance, thereby improving screening performance. For example, Cheng et al. designed a multi-resolution strategy to learn low-resolution patch features, followed by the refinement of high-resolution cell features \cite{cheng2021robust}. Another study encapsulated the statistics of detected cells and then trained a WSI classifier \cite{wang2024artificial}. Additionally, some studies introduced refined segmentation of the nucleus and cytoplasm to characterize cytology morphological semantics \cite{zhu2021hybrid,yu2023ai}.
These methods generally follow a two-step CCS scheme: abnormal cell detection, where high-confidence candidates are identified, and slide-level aggregation for final prediction, illustrated in Extended Data Fig. \ref{F1_intro}(c). While effective on high-quality data \cite{lin2021dual, wang2024artificial}, this CCS scheme faces significant challenges in real-world scenarios. Variations in patient demographics, sample preparation, and staining protocols across institutions often result in inconsistencies between training data and clinical evaluation, leading to a dramatic decline in screening performance \cite{cheng2021robust,zhu2021hybrid}. Addressing these challenges is critical to improve the robustness and generalizability of AI-driven solutions for cancer screening.

Building on the recent success of large-scale pretraining and downstream adaptation, which demonstrates strong generalization and adaptation capabilities \cite{chen2024towards,ma2024towards}, we explore this insight to develop tailored pretraining strategies for computational cytology. 
Firstly, this study proposes the \textbf{Smart-CCS}, a generalizable \textbf{C}ervical \textbf{C}ancer \textbf{S}creening paradigm. 
This paradigm works under a comprehensive strategy that integrates large-scale self-supervised pretraining to capture generalizable feature representations, finetunes the task-specific WSI classification model for cancer screening, and incorporates test-time adaptation to further optimize performance across diverse clinical settings. 
Then, we curated one of the largest multi-center cervical cytology datasets to develop this Smart-CCS system. Finally, the system was retrospectively and prospectively evaluated to demonstrate its clinical applicability and reliability. To the best of our knowledge, Smart-CCS represents the first comprehensive CCS paradigm involving pretraining and adaptation in computational cytology.

%% file: sec/sec2.tex
\subsection*{Large-scale multi-center cervical data collection and annotation}\label{subsec2-1}

\begin{figure*}[htbp]
    \centering
    \includegraphics[width=\linewidth]{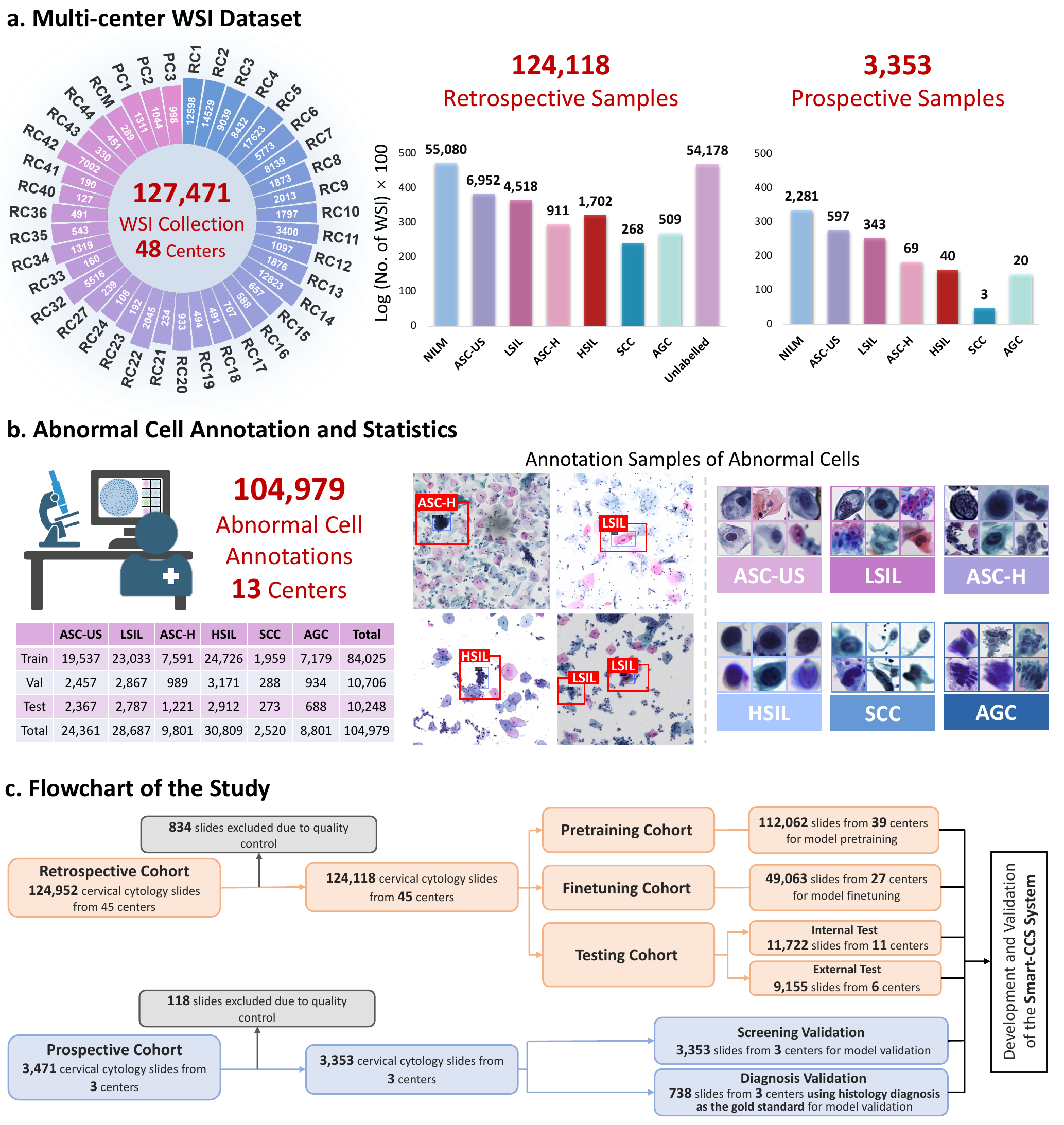}%
    \caption{\textbf{Overview of CCS-127K data and annotations in this study.} \textbf{a}. Class-wise distribution of 127,471 WSIs collected from 48 medical centers, including 124118 samples from 45 retrospective centers and 3,353 samples from 3 prospective centers. Note: RCM represents the centers merged due to limited sample size for each center. \textbf{b}. Abnormal cell annotation statistics: 104,979 abnormal lesion cells were annotated into 6 categories, ASC-US, LSIL, ASC-H, HSIL, SCC, and AGC, termed as CCS-Cell dataset. \textbf{c}. Designed flowchart of the study for the development and validation of the proposed Smart-CCS system. The orange flow represents retrospective studies, while the blue flow represents prospective studies.}
    \label{F2_data}
\end{figure*}
% \clearpage
To develop and validate the Smart-CCS system for generalizable cancer screening, we curated the large-scale and multi-center dataset, named CCS-127K. As illustrated in Fig. \ref{F2_data}(a), we collected a total of 127,471 cytology WSIs from 48 centers. These WSIs were obtained from 128,423 specimens after slide preparation, scanning, and quality control. Details of these processes are elaborated in the Methods section. According to the TBS guidelines \cite{nayar2015bethesda}, cytologists provided slide-level and cell-level annotations. The slide-level labels contained seven major cytology grades including negative for intraepithelial lesion or malignancy (NILM), atypical squamous cells of undetermined significance (ASC-US), low–grade squamous intraepithelial lesion (LSIL), atypical squamous cells cannot exclude an HSIL (ASC-H), high–grade squamous intraepithelial lesions (HSIL), squamous cell carcinoma (SCC), and atypical glandular cells (AGC). In total, cytologists annotated 69,940 retrospective WSIs from 45 centers and 3,353 prospective WSIs from 3 centers.
For cell-level annotations, cytologists delineated regions of interest (RoIs) and annotated abnormal cells using bounding boxes. As shown in Fig. \ref{F2_data}(b), they identified a total of 104,979 abnormal cells from 13 retrospective centers, spanning six abnormal cell types: ASC-US, LSIL, ASC-H, HSIL, SCC, and AGC, named as CCS-Cell dataset. More abbreviations are detailed in Extended Data Table \ref{ST_abb}.

Building on this large and diverse dataset, we designed a structured workflow, illustrated in Fig. \ref{F2_data}(c), that incorporates both retrospective and prospective cohorts. The retrospective cohort was used for model development and evaluation, and it was further divided into three subsets: pretraining, finetuning, and testing. The pretraining cohort, consisting of 112,062 slides from 39 centers, was used for self-supervised learning to initialize the CCS model. This was followed by the finetuning cohort, which included 49,063 labeled slides from 27 centers, to train the model for WSI classification task. For testing, the model was evaluated on a testing cohort divided into two datasets: an internal test dataset with 11,722 slides from 11 centers and an external test dataset with 9,155 samples from 6 independent centers.  Additionally, a prospective cohort of 3,353 slides from three centers was used to evaluate the clinical effectiveness in real-world scenarios. Finally, we collected the 738 corresponding histology diagnoses, which serve as the ground truth to evaluate the cancer diagnosis capability of the Smart-CCS system.

\subsection*{Overview of proposed Smart-CCS system}\label{subsec2-2}
As shown in Extended Data Fig. \ref{SF_method}, we developed a thorough AI-assisted CCS paradigm called Smart-CCS, consisting of three sequential stages: 1) large-scale self-supervised pretraining, 2) CCS model finetuning, and 3) test-time adaptation.

\begin{figure*}[t]
    \centering
    \includegraphics[width=1.0\linewidth]{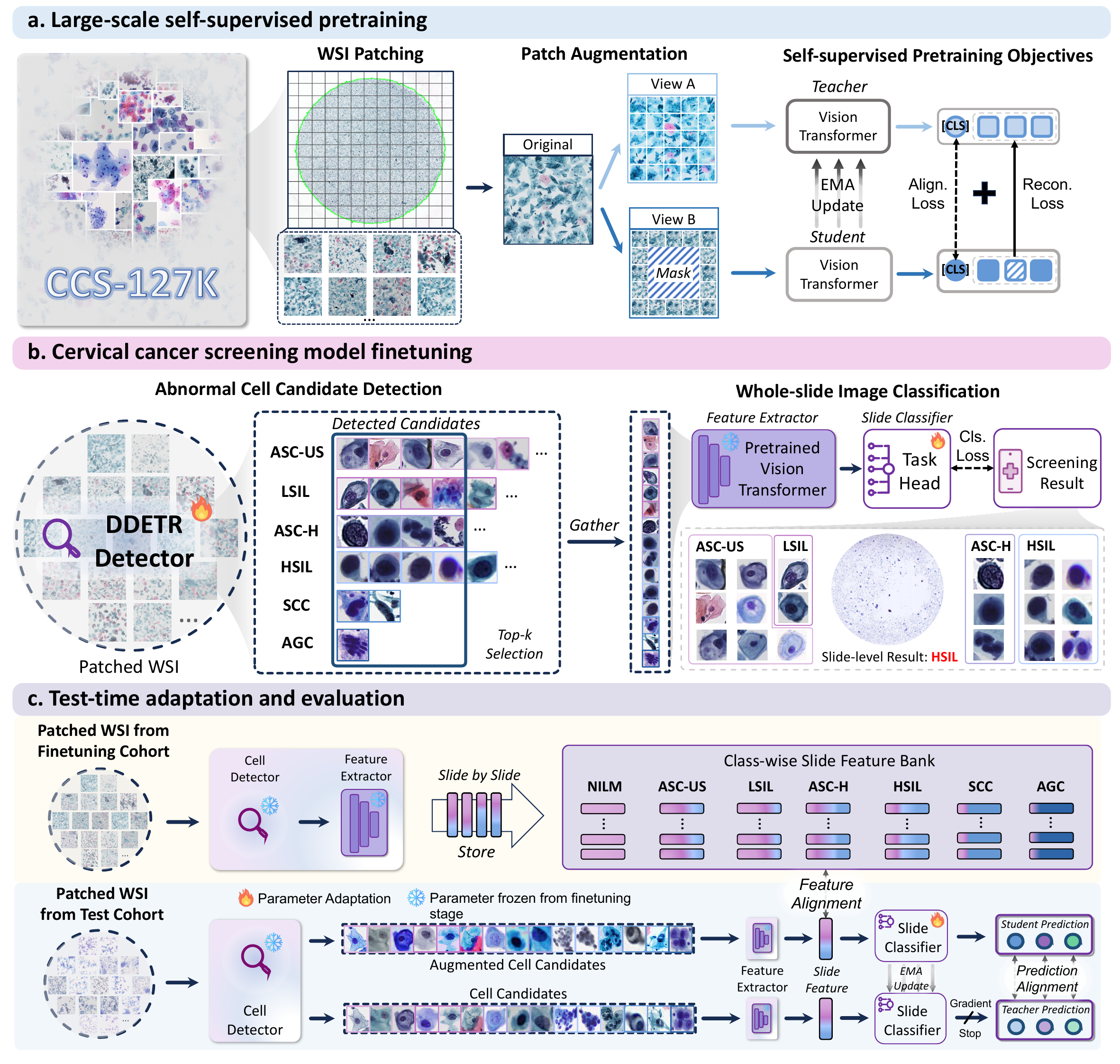}
    \caption{\textbf{Overview of the Smart-CCS Paradigm.} The Smart-CCS paradigm consists of three sequential stages. \textbf{a.} the \textbf{pretraining stage}, which involves large-scale self-supervised pretraining on diverse cytology images from various centers to build a generalizable feature extraction model. \textbf{b.} the \textbf{finetuning stage}, which specializes the pretrained model for cancer screening tasks, including two components: an abnormal cell detector for identifying abnormal cells and a WSI classifier for slide-level  predictions. \textbf{c.} the \textbf{adaptation stage}, which further optimizes trained model for diverse clinical settings via adapting and refining predictions.
    }
    \label{F3_method}
\end{figure*}

The framework of Smart-CCS is illustrated in Fig. \ref{F3_method}. In the pretraining stage, the curated cytology dataset CCS-127K was fully leveraged and exploited in a self-supervised learning manner \cite{oquabdinov2}. Thus, our Smart-CCS could effectively capture and represent inherent and generalizable cytological knowledge such as cellular instance features (cytoplasm, nucleus), semantic features (morphology), and global information (distribution).
In the finetuning stage, the pretrained models were further specialized to CCS tasks under both slide-level and cell-level supervision. Specifically, we followed the two-step CCS scheme involving two models: an abnormal cell detector and a WSI classifier. The detector identified suspicious cells across the entire slide, while the classifier aggregated these cell candidates to generate the final slide classification results. During the adaptation stage, the WSI classification model was optimized to handle unseen samples before making predictions. This approach could enhance the adaptability and generalizability of the Smart-CCS system, enabling it to perform effectively under diverse and complex screening conditions.

%%%%%%%%%%%%%%%%%%%%%%%%%%%%%%%%%%%%%%%%%%%%%

\subsection*{Self-supervised pretraining for cervical cancer screening}\label{subsec2-3}
Large-scale self-supervised pretraining empowers the model with strong generalization capability by yielding robust and off-the-shelf representation in computational cytology. 
In this study, we first investigated the effectiveness of self-supervised pretraining for two crucial tasks in CCS, namely cell-level and WSI-level classification. 

In cell-level tasks, we utilized two public datasets, HErlev \cite{jantzen2005pap} and SIPaKMeD \cite{plissiti2018sipakmed}, along with our collected CCS-Cell dataset to evaluate cell classification performance. We first evaluated scaling law of self-supervised pretraining, the results are illustrated in Fig. \ref{FE1}(a). Overall, the reported cell classification results indicate that pretraining can benefit downstream tasks across all three datasets. Specifically, pretraining using 1 million (M) cytology images yielded top-1 accuracy gains of 4.5\% on HErlev, 1\% on SIPaKMeD, and 3.5\% on CCS-Cell. As scaling pretraining data, classification performance improved steadily and continuously, with increases of 7.2\%, 3.4\%, and 5.6\%, respectively. Ultimately, three datasets reached top-1 accuracies of 0.914 (95\% CI: 0.873–0.955), 0.960 (95\% CI: 0.946–0.974), and 0.883 (95\% CI: 0.868–0.898) when pretrained using 100M cytology patches. 
 To further illustrate the effectiveness, we used t-SNE to visualize cell features with and without pretraining in Extended Data Fig. \ref{SF_tsne}. These visualizations reveal the enhanced feature aggregation capability after pretraining, where cell features are tightly clustered within each category and well-separated from neighboring categories. This aggregation capability potentially addresses category ambiguity issues in classifying cytology grades \cite{jiang2024holistic}.

In WSI-level tasks, we included three retrospective centers consisting of 25,571 WSIs from CCS-127K to investigate the impact of pretraining and investigate the scaling law. We gradually scaled up the pretraining data from 0 to 100M. Shown in Fig. \ref{FE1}(b), the results demonstrated the significant efficacy of pretraining, evidenced by its application in cancer screening through the detection of epithelial cell abnormalities (denoted as ECA), as well as in fine-grained cytology WSI classification tasks (denoted as ALL). As the pretraining data increased, the experimental results steadily improved, showing an overall accuracy increase of 10.34\% in cancer screening and 8.61\% in fine-grained cytology classification, reaching up to 100M data. Ultimately, the three centers achieved 0.950 (95\%CI: 0.942 - 0.959), 0.915 (95\%CI: 0.902-0.929) and 0.958 (95\%CI: 0.946-0.970) accuracies for cancer screening. Details regarding cell and WSI classification are provided in Extended Data Tables. \ref{ST_pretrain_cell}-\ref{ST_pretrain_wsi}.

 Additionally, ablation experiments were conducted on pretraining backbones (ViT-Large, ViT-Gaint) and algorithms (DINOv2 \cite{oquabdinov2}, MoCov3 \cite{chen2021empirical}). Based on the results, we employed the ViT-Large architecture pretrained with DINOv2 using 100 million pretraining data in the following experiments. More experimental results are provided in Extended Data Table. \ref{ST_pretrain_ablation}. 

\subsection*{Retrospective evaluation of abnormal cell detection}\label{subsec2-4}
\begin{figure*}[htbp]
    \centering
    \includegraphics[width=\linewidth]{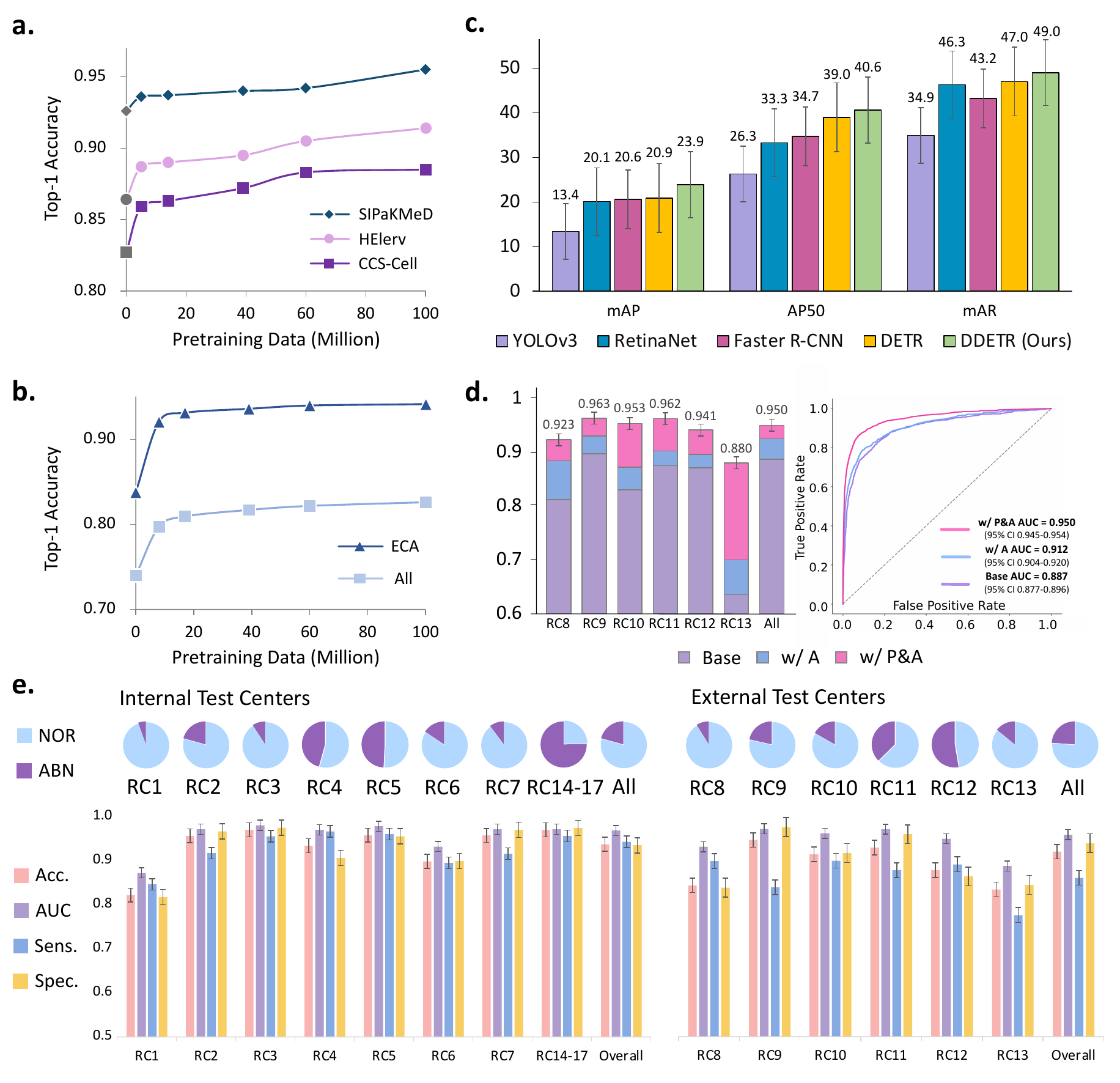}
    \caption{\textbf{Performance of Smart-CCS in retrospective study.} \textbf{a.} Evaluation of cell-level cytology task using cell classification datasets,  SIPaKMeD ($N$ = 4,049), HErlev ($N$ = 918), and CCS-Cell ($N$ = 9,008). \textbf{b.} Evaluation of the WSI-level cytology task using retrospective cervical cytology datasets ($N$ = 5,189, 11,986, 8,396) to assess cancer screening (ECA) and fine-grained classification (ALL) performances. \textbf{c.} Comparison of abnormal cell detection performance among DDETR, DETR, RetinaNet, Faster R-CNN and YOLOv3 on CCS-Cell dataset. \textbf{d.} The external testing performances are evaluated by metric AUC with different settings, Base denotes the typical two-step CCS model, w/ P is introducing pretraining, w/ P\&A refers to our proposed Smart-CCS with pretraining and adaptation. \textbf{e.} Internal and external data distribution, along with the results of cervical cancer screening evaluations.}
    \label{FE1}
\end{figure*}
Abnormal cell detection typically serves as the prerequisite for AI-assisted CCS, where abnormal cells with high confidence scores are identified from the whole slide as candidates and then aggregated for slide-level classification. Thus, we aim to build a strong cell detector to screen out all abnormal and suspicious cells within each WSI. 
As shown in Fig. \ref{FE1}(c), we compared several state-of-the-art (SOTA) detectors used in previous CCS studies, including Faster R-CNN \cite{ren2015faster}, YOLOv3 \cite{zhu2021hybrid,wei2021efficient}, RetinaNet \cite{wang2024artificial}, transformer-based detector DETR \cite{carion2020end}, and its deformable variant DDETR \cite{zhu2020deformable}. 
The experimental results showed that DDETR achieved the best performance and was selected as the detector in Smart-CCS. DDETR surpassed YOLOv3, Faster R-CNN, and RetinaNet by 14.3\%, 5.9\%, and 7.3\% in AP50 (Average Precision with an IoU threshold of 0.50), respectively, benefiting from the powerful visual representation capability of the transformer and the adaptability of deformable attention to multi-resolution \cite{zhu2020deformable}.
The per-class detection results showed that DDETR not only addressed the issue of missed detections for tiny objects, such as SCC (naked nuclei and small nucleoli), achieving a 10.3\% AP50 increase compared to YOLOv3, but also enhanced the detection of morphologically diverse atypical cells. Specifically, it achieved a +13.1\% AP50 improvement in ASC-US and a +8.5\% AP50 improvement in ASC-H (Extended Data Table \ref{ST_det}-\ref{ST_det_p2}). Subsequently, the cell classification performance was evaluated using a confusion matrix, shown in Extended Data Fig. \ref{SF_det}. From this matrix, we observed a clear distinction in the model's ability to identify glandular cells (class AGC) compared to other squamous cell abnormalities. This is attributed to the distinct morphological abnormalities and unique arrangement characteristics of glandular cells. We also found a strong correlation between ASC-US and LSIL, as well as between ASC-H and HSIL, which aligns with clinical practice \cite{nayar2015bethesda,vandenbussche2022cytologic}. Overall, our detector outperforms previous work in differentiating fine-grained cell categories, which serves as the first step in WSI classification model for cancer screening.

\subsection*{Internal testing for multi-center evaluation}\label{subsec2-5}
In the retrospective study, a total of 112,062 cervical cytology samples from pretraining and finetuning cohorts were included for the Smart-CCS system development. Then, internal testing included 11,722 samples from 11 centers for multi-center evaluation of the developed Smart-CCS (Fig. \ref{FE1}(e)). The positive rates among these centers range from 5.48\% 
 to 49.26\% with different grades of epithelial abnormalities.

The quantitative performances in testing centers are presented in Fig. \ref{FE1}(e). 
Overall, Smart-CCS achieved an overall of 0.965 (95\% CI: 0.961–0.969) AUC, 0.965 (95\% CI: 0.961–0.969) accuracy, 0.913 (95\% CI: 0.907–0.919) sensitivity, and  0.896 (95\% CI: 0.889–0.902) specificity on internal testing centers. Specifically, 9 centers achieved AUC metrics greater than  95\% among the 11 internal testing centers, such as 0.971 (95\% CI 0.965–0.978) in RC2, 0.990 (95\% CI: 0.985–0.995) in RC3, 0.985 (95\% CI: 0.977–0.992) in RC5, and 0.971 (95\% CI: 0.960–0.982) in RC7. We also found that centers with a low positive rate, such as RC1, often showed lower classification accuracies. Besides, the system screening capabilities are highly determined by accurately detecting early-stage intraepithelial abnormalities from a large number of samples, as indicated by the metric, sensitivity. Smart-CCS achieved a high overall sensitivity of 0.913 (95\% CI: 0.907–0.919), with 9 internal centers demonstrating sensitivities greater than 85\%, indicating strong screening capabilities.

In terms of different cytology grades, ASC-US+, considered as the squamous cell abnormality, Smart-CCS exhibited high classification capability with an overall AUC of 0.961 (95\% CI: 0.957–0.965) and a sensitivity of 0.910 (95\% CI: 0.904–0.916). Then, the critical cytology group LSIL+, to be considered for further colposcopy, maintained advanced screening performance of 0.958 (95\% CI: 0.953–0.962) AUC with 0.910 (95\% CI: 0.903–0.916) sensitivity. Especially in RC1, the challenging center with a quite low positive rate (5.48\%), Smart-CCS increased from 0.767 (95\% CI: 0.750–0.784) in ECA to 0.902 (95\% CI: 0.890–0.914) in LSIL+, demonstrating the effect of ASC-US, the ambiguous category \cite{jiang2024holistic, nayar2015bethesda}. The HSIL+ group typically refers to malignant neoplasia, with superior classification results ranging from 0.918 (95\% CI: 0.904–0.931) to 0.994 (95\% CI: 0.987–1.001) in AUC within internal testing centers. Furthermore, we also reported the F1 score metric to assess the model stability under class imbalance. Smart-CCS achieved favorable F1 scores of 0.903 (95\% CI: 0.896–0.909), 0.897 (95\% CI: 0.890–0.903), 0.891 (95\% CI: 0.884–0.897), and 0.950 (95\% CI: 0.945–0.955) in subgroups ECA, ASC-US+, LSIL+, and HSIL+. More result details are appended in Extended Data Tables. \ref{ST_wsi}-\ref{ST_wsi_3}.

\subsection*{External testing for generalizability evaluation}\label{subsec2-5}
The cytology samples collected from different centers present diverse and large variations, which brings challenges for system generalizability. Under the proposed Smart-CCS paradigm, the pretrained model could extract general and strong cytology features that are domain-invariant across diverse clinical centers.  To evaluate the effectiveness and generalizability of developed Smart-CCS, we retrospectively included 9,155 cytology WSIs from six independent centers (RC8-RC13).

In the external testing, the positive rates among six centers ranged from 8.81\% (RC8) to 52.65\% (RC12), resulting in an overall rate of 23.89\%. The results are shown in Fig. \ref{FE1}(e), where we obtained an overall screening performance of 0.912 (95\% CI: 0.906–0.918) accuracy, 0.950 (95\% CI: 0.945–0.954) AUC, and 0.854 (95\% CI: 0.847–0.861) sensitivity. Specifically, Smart-CCS achieved consistent and favorable classification results ranging from 0.880 (95\% CI: 0.859–0.901) to 0.963 (95\% CI: 0.954–0.972) AUC across six external centers.
The external testing performance reveals the strong generalization capabilities of Smart-CCS.

To further investigate the efficacy of the proposed pretraining and adaptation strategies in Smart-CCS, we conducted ablation studies comparing three experimental groups: a two-step CCS model (denoted as Base), CCS with pretraining (denoted as w/ P), and our Smart-CCS with both pretraining and adaptation (denoted as w/ P\&A). As illustrated in Fig. \ref{FE1}(d), we report the area under the receiver operating characteristic curve (AUC) for cancer screening across different centers, as well as the overall performance. Compared to the CCS baseline model, pretraining consistently increased the AUC, from 0.897 (95\% CI: 0.877–0.918) to 0.930 (95\% CI: 0.913–0.948) in RC9, and from 0.830 (95\% CI: 0.805–0.855) to 0.911 (95\% CI: 0.892–0.930) in RC10, among others. Furthermore, when both pretraining and adaptation were applied, Smart-CCS achieved a 6.3\% AUC improvement against the CCS baseline model. The demonstrated effectiveness of the Smart-CCS paradigm in external testing underscores its potential clinical applicability in complex scenarios. Additional details regarding the ablation studies can be found in Extended Data Table \ref{ST_ext}. Moreover, we compared classifiers, including MeanMIL, MaxMIL, ABMIL \cite{ilse2018attention}, DSMIL \cite{li2021dual}, CLAM \cite{shao2021transmil}, TransMIL \cite{lu2021data}, and S4MIL \cite{fillioux2023structured}, in terms of internal and external testing performance, as shown in Extended Data Table \ref{ST_clas}.

\subsection*{Prospective study for clinical validation}\label{subsec2-6}
\begin{figure*}[htbp]
    \centering
    \includegraphics[width=1.0\linewidth]{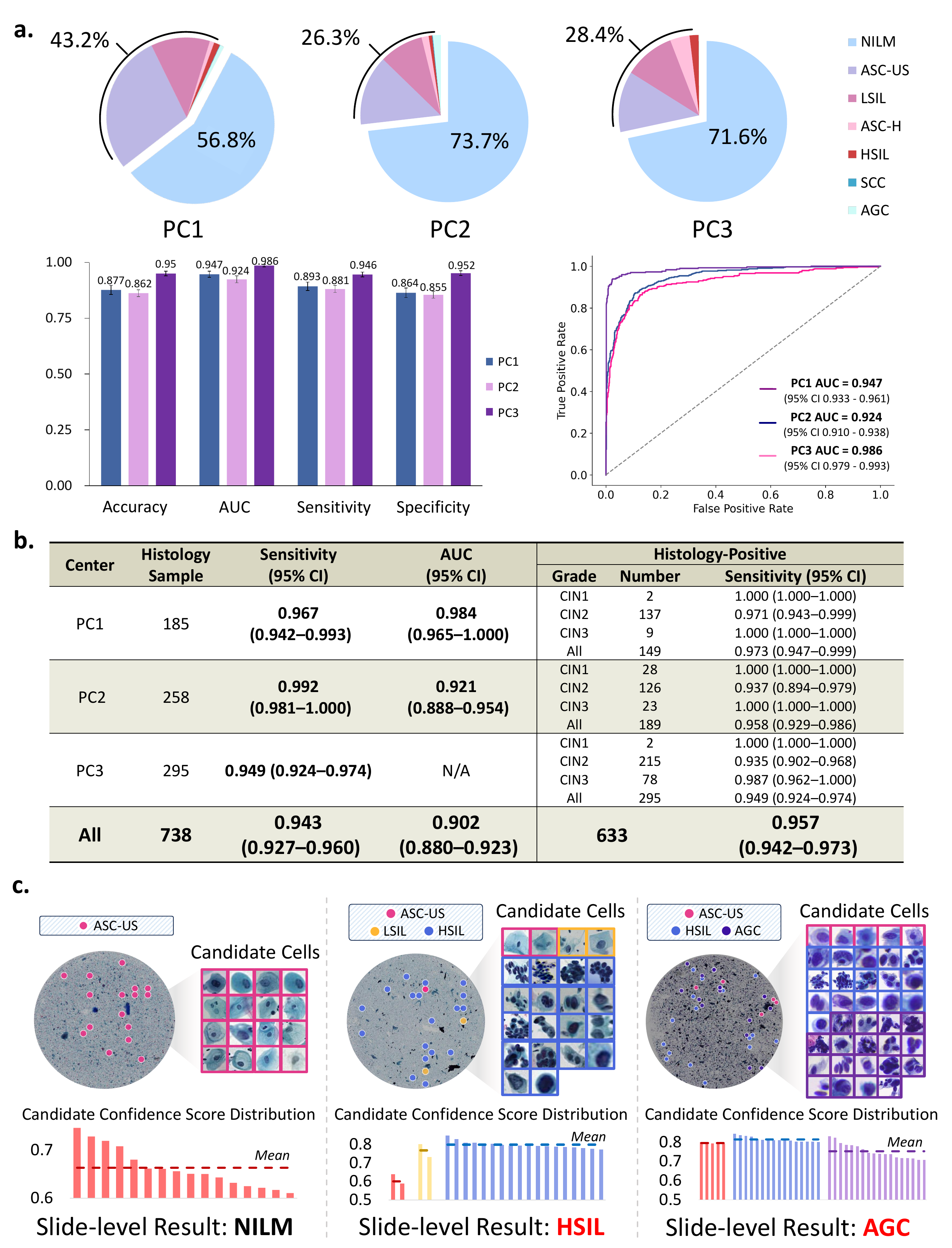}
    \caption{\textbf{Performance of Smart-CCS in the prospective study.} \textbf{a}. Grade distribution among three prospective centers (PC1, PC2 and PC3) and reported evaluation results from our Smart-CCS system. \textbf{b}. The diagnostic performance of Smart-CCS in cancer detection using histological results as the gold standard. \textbf{c}. Visualizations from cervical cancer screenings, which include interpretable results at both the cell-level and slide-level, derived from three sample tests.}
    \label{FE2}
\end{figure*}

Between January 1 2024 and July 31, 2024, we recruited a total of 3,353 participants from three prospective centers, PC1, PC2 and PC3 to form the prospective cohort. 
As shown in Fig. \ref{FE2}(a), three prospective centers individually contributed 998, 1,311, and 1,044 samples, with positive rates of 43.24\%, 26.32\%, and 28.35\%. Besides, we also obtained 185, 258, and 295 corresponding histological diagnosis results as the gold standard for further system evaluation.

%%%%%%%%%%%%%%%%%%%%%%%%%%%%%%%%%%%%%%%%%%%%%

Overall, the evaluation results of Smart-CCS for cancer screening demonstrated consistent generalization capabilities in prospective centers, as shown in Fig. \ref{FE2}(a) and Extended Data Table. \ref{ST_pros}. Specifically, Smart-CCS achieved accuracies of  0.877 (95\% CI: 0.856–0.897), 0.862 (95\% CI: 0.843–0.881), and 0.950 (95\% CI: 0.937–0.963) across three prospective centers, with high sensitivities of 0.893 (95\% CI: 0.874–0.912), 0.881 (95\% CI: 0.864–0.899), and 0.946 (95\% CI: 0.932–0.960). The high AUC scores ranging from 0.924 (95\% CI: 0.910–0.938) to 0.986 (95\% CI: 0.979–0.993) also reveal the robust and strong cancer screening capability of Smart-CCS system in clinical scenarios.

To further validate Smart-CCS for cervical cancer diagnosis, we utilized histological biopsy results as the gold standard \cite{ouh2021discrepancy}. Histology provides a more confirmative diagnosis through direct tissue examination compared to the individual cell analysis of cytology, making it a reliable benchmark for assessing the accuracy of the cytology-based Smart-CCS. The overall and different grades (i.e., CIN1, CIN2, CIN3) results are presented in Fig. \ref{FE2}(b). Overall, Smart-CCS obtained a sensitivity of 0.943 (95\% CI 0.927-0.960) and an AUC of 0.902 (95\% CI 0.880-0.923) in cervical cancer diagnosis. Specifically, the three prospective centers reported sensitivities of 0.967 (95\% CI: 0.942–0.993), 0.992 (95\% CI: 0.981–1.000), 0.949 (95\% CI: 0.924–0.974), with corresponding AUCs of 0.984 (95\% CI: 0.965–1.000), 0.921(95\% CI: 0.888–0.954), and N/A due to the lack of negative samples in PC3. For histology-positive samples, Smart-CCS achieved an overall of 0.957 (95\% CI: 0.942–0.973) sensitivity in three centers. The high sensitivities can be observed across all groups, ranging from 0.935 (95\% CI: 0.902–0.968) to 1.000 (95\% CI: 1.000–1.000). We also provided the results of different histology grades, CIN1, CIN2, and CIN3. We found there were no missed samples in CIN1 and CIN3 in two centers (PC1, PC3). The CIN2 group achieved 0.971 (95\% CI: 0.943–0.999), 0.937 (95\% CI: 0.894–0.979) and 0.935 (95\% CI: 0.902–0.968) sensitivities.
Notably, Smart-CCS yielded sensitivities close to 1.00 for HSIL+ across all three prospective centers, demonstrating a high consistency between cancer screening and diagnosis on higher cytology grades. 
Therefore, generalizable cancer screening could potentially avoid unnecessary biopsies and colposcopies for patients at risk of cervical cancer. 

\subsection*{Interpretability analysis}
To support clinical practice, we illustrate the decision-making process of Smart-CCS in Fig. \ref{FE2}(c), providing interpretable insights from diverse perspectives.
 At the slide-level, Smart-CCS predicts positive scores for precancerous abnormalities based on cervical cytology specimens. It provides detailed prediction categories and scores for malignancy grade evaluation, which inform subsequent colposcopy and biopsy procedures. For instance, in Fig. \ref{FE2}(c), the left WSI sample achieves a confidence score of 0.9999 for the NILM category, while the middle sample yields a positive score of 0.9999, with a high-grade HSIL confidence of 0.9915. This indicates a high probability of cervical neoplasia and malignancy.

In the context of cell-level screening, suspicious cells are highlighted using distinct markers in each WSI. The visualization illustrates that these suspicious cells are randomly distributed across the WSIs due to the uniform mixing and centrifugation during specimen preparation \cite{jiang2023deep}. Furthermore, detected suspicious cells, along with confidence score statistics and characteristic morphologies, provide reliable clues for the final diagnosis. In the middle sample illustrated in Fig. \ref{FE2}(c), Smart-CCS detected a few  LSIL cells (i.e., koilocytotic cells with typical perinuclear halos) alongside a significant number of HSIL cells, which demonstrated markedly enlarged nuclei, reduced cytoplasmic areas, and hyperchromatic clustering. This predicted information aligns well with the ground truth for this sample, which is HSIL. Similarly, the AGC sample reveals a substantial presence of AGC cells arranged in a fence-like pattern, alongside other atypical cells, as shown in Fig. \ref{FE2}(c)(right). In the case of NILM, despite a slide-level prediction score of nearly 1.00, some cells exhibit deepening and enlargement of nuclei, presenting as non-neoplastic cellular variations or hard mimics \cite{nayar2015bethesda}. By integrating these insights within an interactive interface, we have developed Smart-CCS into a comprehensive system, as depicted in Extended Data Fig. \ref{SF_integrate}. This system aims to provide cytologists with interpretable AI-assisted results, ensuring reliable screening outcomes and guiding subsequent final diagnoses, thereby enhancing the overall diagnostic workflow.

%% file: sec/sec3.tex
Cervical cancer has been targeted for global elimination under the WHO initiative \cite{world2020global}. Leveraging AI to assist cytologists can significantly accelerate cervical cancer screening, especially in large-scale precancerous screening conditions. In recent years, several AI-based computational cytology studies preliminarily explored its feasibility, particularly a recent study demonstrated the effectiveness of AI-assisted cytology in cancer identification and tumor origin prediction, highlighting the potential of AI in computational cytology \cite{tian2024prediction,rassy2024predicting, li2024new}. However, previous AI-assisted CCS systems often faced challenges in generalization and robustness across different clinical scenarios, particularly concerning variations in slide preparation and imaging protocols \cite{cheng2021robust,wang2024artificial,zhu2021hybrid,wu2024development}. 

In this study, we introduced Smart-CCS, a generalizable cervical cancer screening paradigm. This pioneering paradigm consists of three stages: self-supervised pretraining, screening model finetuning, and test-time adaptation. It leverages: 1) large-scale self-supervised pretraining for robust cytology representations, 2) effective utilization of both cell-level and slide-level supervision, and 3) model adaptation during clinical evaluation. To support this, we constructed a large, multi-center dataset, CCS-127K, which includes 127,471 cervical cytology WSIs from 48 centers. To the best of our knowledge, this is the first comprehensive paradigm for cervical cancer screening with multi-center and prospective validation.

The experimental results of internal testing, external testing, and prospective studies revealed the extensive effectiveness and potential clinical benefits of the proposed Smart-CCS for generalizing cancer screening. 
Notably, the significant improvements in cell-level and WSI-level downstream tasks demonstrated the efficacy of pretraining in capturing general cytology information. For retrospective evaluation of abnormal cell detection, we compared SOTA detection models and selected the best as the first step in WSI classification model. In terms of internal testing, Smart-CCS achieved an overall AUC of 0.965 (95\% CI: 0.961–0.969) across 11 centers for cancer screening. 
In external testing, Smart-CCS achieved an AUC from 0.880 (95\% CI: 0.859–0.901) to 0.941 (95\% CI: 0.927–0.955)  across 6 external centers with high sensitivities, validating its applicability in diverse clinical scenarios. Additionally, prospective studies in PC1-PC3 yielded high AUC scores of 0.947 (95\% CI: 0.933–0.961), 0.924 (95\% CI: 0.910–0.938), and 0.986 (95\% CI: 0.979–0.993), respectively. Further histological evaluation and interpretability analysis yielded consistent conclusions with clinical knowledge, which highlights the reliability and superiority of the proposed Smart-CCS paradigm.

This study has several limitations. 
First, the availability of large-scale data is critical for pretraining robust and generalizable models. Although we have constructed one of the largest cytology datasets to date, comprising 127,471 WSIs, it remains relatively small compared to datasets commonly used in the histology domain. For instance, Virchow2 was pretrained on 3.1M WSIs \cite{vorontsov2024foundation}, and CONCH utilized 1.17M image-caption pairs \cite{lu2024visual}, leveraging extensive public histology datasets such as TCGA \cite{weinstein2013cancer}. Similarly, recent advancements in cytology have also focused on scaling up data resources to enhance model development \cite{yu2023ai}. To address this limitation, we are actively working to incorporate larger datasets for both pretraining and validation, with the goal of advancing cancer screening and computational cytology.
Second, from a methodological perspective, the ability to distinguish individual cell instances is critical for guiding accurate screening decisions. While we have developed a detection-based WSI classification framework, integrating fine-grained cellular information could further improve the modeling of cell-to-WSI relationships. For example, transitioning from patch-level pretraining to cell-level pretraining or incorporating quantified morphological features, such as cytoplasmic segmentation masks and nucleus-to-cytoplasm ratios, may enhance the performance and interpretability of the Smart-CCS system \cite{zhu2021hybrid}.
Third, regarding system validation, although this study included independent testing across 11 external centers and prospective evaluation in 3 centers, broader deployment and validation in diverse clinical settings are essential to facilitate widespread adoption. Future evaluations should consider a wider range of demographic variables, including ethnicity, age, geographic regions, and variations in slide preparation and imaging protocols, to ensure the system's generalizability and robustness across heterogeneous clinical scenarios.

Ultimately, to establish a trustworthy cancer screening system for large-scale clinical applications, several in-depth exploration directions follow this study. First, establishing a unified cytology WSI benchmark is essential to tackle unique challenges such as cell instance distinguishability and ambiguous categorization of atypical findings (e.g., ASC-US, ASC-H) \cite{jiang2024holistic}.  
Secondly, while Smart-CCS demonstrated the effectiveness for cervical cancer screening, the paradigm has significant potential for broader applications. Generalizing Smart-CCS to other cancers, such as urine \cite{wu2024development}, thyroid \cite{wang2024deep}, and pleural effusion samples \cite{tian2024prediction}, could contribute to a cytology foundation model, enhancing its applicability in computational cytology. Finally, while cytology primarily focuses on cancer screening, large-scale data-driven approaches could expand to more clinical applications, including cancer diagnosis \cite{wu2024development}, survival prognosis \cite{sigel2017cytology,valletti2021gastric}, biomarker prediction \cite{tian2024prediction,rassy2024predicting,li2024new} and molecular-level discovery \cite{caputo2024current, ohori2024molecular}.

In summary, the Smart-CCS paradigm demonstrates strong potential for advancing cervical cancer screening while paving the way for broader applications in computational cytology.

%% file: sec/sec4.tex
\subsection*{Data collection and annotation}\label{subsec4-1}
The established CCS-127K dataset is a highly diverse collection of cervical cytology specimens, enriched with both cell and slide annotations. In total, we compiled 128,423 cytological specimens from 48 different centers. The data collection process involved several key steps. First, the specimens were prepared using commonly employed cytology sedimentation methods, including natural, membrane, and centrifugal techniques. Next, all liquid-based cytology specimens were digitized into WSIs using various imaging protocols, which contributed to the diversity of cytology images within the CCS-127K dataset. Four distinct scanners with varying specifications were employed: the Pannoramic MIDI (3DHISTECH Ltd.), SQSL-510 (Shenzhen Shengqiang Technology Ltd.), Aperio AT2 (Leica Microsystems Ltd.), and HDS-MS-200A (Xiamen Heidstar Ltd.). 
Subsequently, quality control procedures were conducted, resulting in a final tally of 127,471 specimens available for development and evaluation within the Smart-CCS system. During data annotation, cytologists provided both cell-level and WSI-level annotations according to TBS guidelines \cite{nayar2015bethesda}. For cell-level annotations, cytologists first identified RoIs containing abnormal cells and then utilized bounding boxes to delineate all abnormal cells within the RoI. These annotations consisted of six major categories of abnormalities: ASC-US, LSIL, ASC-H, HSIL, SCC, and AGC,  and were used to develop the abnormal cell detector. It is noteworthy that the AGC category encompasses typical glandular cells exhibiting epithelial cell abnormalities, including AGC-NOS, AGC-FN, AIS, and ADC. This study specifically focuses on cancer screening, and consequently, organism categories from TBS were excluded from further analysis. For slide-level annotations, we obtained corresponding diagnostic results from each participating center. These results were assessed by two cytologists, each possessing over ten years of experience, who examined the cytology specimens using a microscope. In case of discrepancies, a third senior cytologist, with over fifteen years of experience, rendered the final decision. The slide-level labels comprised seven cytology grades: NILM, ASC-US, LSIL, ASC-H, HSIL, SCC, and AGC. Additionally, we gathered 738 corresponding histological diagnostic results from prospective cohorts to further evaluate the system's diagnostic capabilities. When processing biopsy results, benign and inflammatory findings were classified as negative, while carcinoma cases were categorized as CIN3 for diagnostic evaluation.

\subsection*{Data preprocessing}\label{subsec4-2}
Each cervical digital slide typically reaches a giga-pixel resolution of up to 100,000 $\times$ 100,000 pixels, which needs to be tiled into patches for model input to fulfill the computation needs, shown in Extended Data Fig. \ref{SF_preprocessing}. 
During slide preprocessing, we employed OpenSlide library \cite{goode2013openslide} and CLAM toolbox \cite{lu2021data} for foreground extraction and slide tiling. The CLAM toolbox first used thresholding segmentation to remove a significant amount of irrelevant background, and then performed non-overlapping patching at \(1,200 \times 1,200\) pixels in the foreground. The final statistics of WSI and patch are summarized in Extended Data Table. \ref{ST_patches}.

\subsection*{Quality control for slides and patches}\label{subsec4-3}
Strict quality control is essential for both cytologist screening and AI system development. There are several factors contributing to the reduced quality, including sample preparation, staining reagents, and scanning protocols, which may impede the effective representation of deep features. In clinical practice, cytologists often refer to TBS criteria for sample adequacy assessment \cite{nayar2015bethesda}, including minimum squamous cellularity criteria, transformation zone component, obscuring factors, and interfering substances, along with imaging. Following this, we established the exclusion criteria, shown in Extended Data Fig. \ref{SF_qualitycontrol}, considering a) preparation (impurities, dried specimen, interfering labeling), b) staining (overstaining, uneven staining), c) scanning (out-of-focus, severe artifacts), d) minimum cellularity. We additionally employed a threshold-based foreground segmentation algorithm to assess cytology patch quality, effectively excluding those with insufficient foreground proportion. As a result, we excluded 952 WSIs and 46,136 patches from the CCS-127K dataset to ensure high-quality cytology data for system development and evaluation. 

\subsection*{Large-scale self-supervised cytology pretraining}\label{subsec4-3}
Large-scale self-supervised pretraining has made significant advancements in computational pathology, benefiting greatly from large-scale publicly available histology datasets \cite{huang2023visual,lu2024visual}. For instance, CtransPath was pretrained using 32K WSIs (4.2M patches) from public TCGA and PAIP datasets \cite{wang2022transformer}, and UNI was pretrained by 100K WSIs (100M patches) collected from private data and the GTEx consortium \cite{chen2024towards}.
This resulted in the development of generalizable pretrained models that promote downstream histological tasks. However, the inaccessibility of cytological data significantly hinders large-scale pretraining in computational cytology. To overcome this limitation, we collected more than 127K cytology WSIs with 227 million patches to support large-scale pretraining in Smart-CCS.

 The structure of visual pretraining is provided in Fig. \ref{F3_method}(a), we utilized the SOTA self-supervised learning framework, DINOv2 \cite{oquabdinov2}. This framework employs a student-teacher network, where the student model is supervised by pseudo labels from the teacher model. Thus, the network output distributions are aligned for knowledge distillation. This pretraining framework relies on alignment strategies, primarily achieved through reconstruction loss and alignment loss between the outputs of the student and teacher networks. Specifically, given a batch of cytology patches, they are first randomly augmented, including color jittering, gaussian blur, and solarization, to generate multiple augmented patches, which are then subjected to global crop and multiple local crops. The global crop patches are randomly masked for masked image modeling.
Then, the global crops without masks serve as inputs to the teacher network, while the global crops with masks and local crops serve as inputs to the student network, outputting crop prototypes. Subsequently, prototype alignment involves reconstruction loss and alignment loss to update the student network. Masked image modeling in reconstruction loss utilizes the student's output to predict the teacher's output, encouraging the student network to learn semantic and contextual information based on the surrounding pixels of the masked regions. Regarding alignment loss, aggregated crop prototypes (both global crop with masks and local crop) from the student are aligned with the corresponding global crops without masks through minimizing cross-entropy loss. Finally, the teacher network is updated by the exponential moving average of the student network. More pretraining details and settings are presented in the Extended Data Table. \ref{ST_pretrain_para}.

Leveraging such self-supervised pretraining, the generalizable cytological knowledge can be captured and accumulated from large-scale diverse data for general cytology understanding. After pretraining, the pertained model serves as a feature extractor for downstream cancer screening tasks. In that case, the extracted features can efficiently generalize and adapt to task-specific whole slide image classification tasks equipped with supervision information, resulting in strong and consistent performance in multi-center retrospective and prospective validations.

\subsection*{Cytology WSI classification model finetuning}\label{subsec4-4}
The objective of this study is to develop a trustworthy and generalizable WSI classification model for cervical cancer screening. 
Existing cytology WSI classification studies have consistently adhered to the two-step WSI classification approach, specifically focusing on abnormal cell detection and slide-level aggregation \cite{lin2021dual,wang2024artificial,zhu2021hybrid,yu2023ai}. This aligns with the clinical setting where cytologists first identify suspicious cells and then provide patient diagnostic results. 
These studies are consistent in the first step, developing a strong abnormal cell detector for atypical or malignant cell detection, while different in the second step with diverse designs of cell aggregation. Some studies introduced morphological segmentation of atypical cells (e.g., ASC-US) for refined guidance towards final diagnosis \cite{zhu2021hybrid,yu2023ai}. However, pixel-level annotation of cytoplasm and nuclei entails significant labor burden for system development. 
Other designs involving cell statistics or rule-based WSI classifiers resulted in screening performance highly dependent on detector performance, as weak classifiers limit generalization capability \cite{lin2021dual,wang2024artificial}. Additionally, permutations and combinations of models can increase the model complexities \cite{zhu2021hybrid,wu2024development}. Recently, another stream followed histology researches and adopted multi-instance learning (MIL) scheme for WSI classification \cite{zhao2024less,jin2024hmil}. 
This stream ignores cell information and just relies on slide-level supervision. Therefore, leveraging both cell-level supervision from detection and slide-level supervision from MIL is the main consideration in algorithm design. 

The proposed cancer screening algorithm is provided in Fig. \ref{FE1}(b), consisting of two main networks, i.e., abnormal cell detection and slide-level classification. Given a cytology WSI, local patches are extracted from the WSI using a non-overlapping sliding window of 1,200$\times$1,200 pixels. 
These patches are then inferred by the selected DDETR \cite{zhu2020deformable} to identify abnormal cells as abnormality candidates with predicted information including locations, types, and confidences. The first step primarily suppresses diagnosis-irrelevant information such as background and normal cells, while the second stage aggregates informative candidates and learns diagnosis-related representations. In the second step, after sliding through the entire WSI, a bag is constructed by aggregating selected cell images, which are cropped from WSI based on the predicted locations, classes and confidences. Subsequently, predicted abnormal cell images in the WSI bag are mapped into generalizable slide features through the frozen feature encoder from the pretraining stage and propagated through the classification head to obtain the final diagnosis results. 
This cytology WSI classification algorithm provides both abnormal cells and suggested diagnostic results. More techniques about the two steps are detailed as follows.

\noindent \textbf{Abnormal cell detection.} In abnormal cell detection, the cytology images are first augmented and then mapped into multi-scale feature maps through the backbone with 1/8, 1/16, 1/32, 1/64 of the original size. The multi-scale feature maps serve as the input to the transformer structure, which includes encoder and decoder layers. Each encoder layer comprises the following components, positional encoding, residual connection, layer normalization, feedforward neural network \cite{dosovitskiy2020image}, and the proposed multi-scale deformable attention module (MSDAttn). This module, designed for capturing deformable information, is a variant of multi-head self-attention \cite{dosovitskiy2020image} and implemented as follows,
\begin{equation}
\operatorname{MSDAttn}\left(\boldsymbol{z}_q, \hat{\boldsymbol{p}}_q,\left\{\boldsymbol{x}^l\right\}_{l=1}^L\right)=\sum_{m=1}^M \boldsymbol{W}_m\left[\sum_{l=1}^L \sum_{k=1}^K A_{m l q k} \cdot \boldsymbol{W}_m^{\prime} \boldsymbol{x}^l\left(\phi_l\left(\hat{\boldsymbol{p}}_q\right)+\Delta \boldsymbol{p}_{m l q k}\right)\right]
\end{equation}
where $M$, $L$, and $K$ are the number of attention head, feature level, and sampling point, respectively. $\boldsymbol{z}_q$ and $\hat{\boldsymbol{p}}_q$ denote content feature and the normalized coordinates of the reference point for each query element $q$, and $\boldsymbol{x}^l$ are the multi-scale feature maps.  $A_{m l q k}$ and $\Delta \boldsymbol{p}_{m l q k}$ are the attention weight and the sampling offset projected from $\boldsymbol{z}_q$. $\phi_l\left( \cdot \right)$ represent a scaling function. $\boldsymbol{W}_m$ and $\boldsymbol{W}_m^{\prime}$ are learnable weights. 
In the decoder, there are the same components with the object queries input to constrain the maximum number of detections. The final prediction head, a feed-forward network acts as the classification and regression head, predicting the probability and bounding box coordinates. The above multi-scale deformable design is highly aligned with the challenges of cancer screening, 1) small object detection, such as naked nuclei cells, 2) multi-scale for multi-resolution cytology images, 3) diverse morphological characteristics. 

In this study, we also evaluated other SOTA object detection models, namely Faster R-CNN \cite{ren2015faster}, YOLO-v3 \cite{redmon2018yolov3}, RetinaNet \cite{lin2017focal}, DETR \cite{carion2020end}. In the cancer screening algorithm, we first train the abnormal cell detector using the cell-level CCS-Cell dataset with instance annotations, and then adopt the trained detector to screen abnormal cells in the first step of CCS algorithm.

\noindent \textbf{Whole slide image classification.} In the second step, WSI aggregation, thousands of detected abnormal cells with each WSI are aggregated for the final WSI classification.
Regarding cell selection and aggregation, we consider the top-k confident cell images for each abnormal class, ASC-US, LSIL, ASC-H, HSIL, SCC and AGC, generating an WSI feature bag to represent original WSI. Subsequently, each cell image in the bag is augmented as input for the classifier. 

 In terms of WSI classifier, cell images are first mapped into 1024-dimensional features by our pretrained vision transformer with frozen weights. All mapped instance features in the WSI bag are concatenated to form a feature bag with size of $n$×1024 (where $n$ is the number of instances), serving as the generalized representation of the original WSI. 
 Finally, the feature bag of abnormal cells passes through trainable classification heads to obtain slide-level predictions. In the ablation experiments, we compared several SOTA classification heads, including MIL-based models such as MeanMIL, MaxMIL, ABMIL \cite{ilse2018attention}, CLAM \cite{lu2021data}, DSMIL \cite{li2021dual}, TransMIL \cite{shao2021transmil} and S4MIL \cite{fillioux2023structured}. 

\subsection*{Test-time adaptation with prototype alignment}\label{subsec4-7pro}
To tackle the complex clinical settings, we introduced adaptation techniques into our Smart-CCS diagram to adapt the developed model in the inference stage, further enhancing system robustness.
The satisfactory performance of deep learning models relies on a strong assumption that the training and testing data are drawn from the same or similar distribution, namely independently and identically distributed (IID) \cite{quinonero2022dataset}. However, for AI systems, cross-center data variability poses a significant challenge in clinical deployment, often leading to severe performance degradation  \cite{asif2021towards}. This issue primarily stems from variations in specimen preparation and imaging protocols across different hospitals, which can cause shifts in cytology WSI data. These shifts may manifest as differences in resolution, staining intensity, or overall image appearance, potentially introducing biases toward specific staining and imaging protocols.

Test-time adaptation considers the settings involving adapting a trained task-specific model to the current test batch before making predictions. Thus, we introduce the test-time adaptation into Smart-CCS to optimize the trained model by dynamically updating model weights during inference. To be more concrete, we propose a test-time adaptation approach with prototype alignment, which enhances the adaptability of the trained WSI classification model by leveraging both the current test batch and source supervision information. This adaptation strategy ultimately improves the model's performance in real-world clinical scenarios.
Firstly, we extract WSI features using pretrained extractor from retrospective data, with size of $n$×1024 (where $n$ is the number of instances).
The top-$k$ confident WSI features for each category constitute a designed class-wise prototype bank that stores the diagnostic supervision information from the retrospective samples. 
Subsequently, for WSIs from current clinical scenarios, the model infers abnormal cells as candidates. 
For adaptation, given the current batch $x$ with $m$ samples, the $i_{th}$ sample containing $n$ cell instances is augmented to improve the diversity of cell instances \cite{wang2022continual}. The current batch $x$ with its augmented view $\tilde{x}$ are transformed into features $f$ and $\tilde{f}$. Then, we build a mean teacher framework to distill the alignment knowledge from source to current test samples. The teacher and student are initialized by the trained slide classifier. Subsequently, the knowledge is distilled through feature alignment loss and prediction alignment loss. The feature alignment loss aligns current batch $f$ with augmented view $\tilde{f}$ and prototypes via contrastive learning. Here, the positive pairs in $J^{+}$ are built by current sample $I$ with corresponding augmentation, and the nearest prototypes computed by the cosine similarity. Then, the feature alignment loss is formed as, 
\begin{equation}
\mathcal{L}_{Falign}= - \sum_{i \in I}  \sum_{j^{+} \in J^{+}} \log \frac{\exp \left(\frac{Sim(z_i, z_{j^{+}}) }{ \tau}\right)}{\sum_{j \in J} \exp \left(\frac{Sim(z_i \cdot z_j)}{ \tau}\right)},
 \end{equation}
where $Sim(z_{i},z_{j})=z_{i} \cdot z_{j} /(\left\|z_{i}\right\|\left\|z_{j}\right\|)$.

$z_{i}$ represents the $i_{th}$ embedding after projection, and $\tau$ is a tunable temperature hyper-parameter.
For the denominator, $J=J^{+}+J^{-}$ denotes the total number of positive ($J^{+}$) and negative ($J^{-}$) samples. In terms of the prediction alignment loss, student and teacher networks output predictions $\tilde{y}^{s}_{ic}$ and $\tilde{y}^{t}_{ic}$ for given inputs $x_{i}$. Then, the consistency loss $\mathcal{L}_{Palign}$ is employed to align these predictions, formed as,
 \begin{equation}
\mathcal{L}_{Palign}=  -\sum_{c=1}^C \tilde{y}^{s}_{ic} \log \tilde{y}^{t}_{ic}
 \end{equation}
where $C$ is the total number of classes.
Finally, the student network is updated using the overall loss. The teacher network is updated stably using an exponential moving average. To keep prediction smooth, the ensemble of the teacher output and the student output serves as the final prediction after test-time adaptation.

In conclusion, pretraining leverages general cytology information, finetuning specializes the pretrained model for cancer screening specific tasks, and adaptation further optimizes the finetuned model for clinical evaluation settings. Together, these three techniques within Smart-CCS significantly advance the generalizability of cervical cancer screening.

\subsection*{Model training and implementation}\label{subsec4-6}
In the experiment, we utilized Python (v3.9.0) and PyTorch (v2.0.0) \cite{paszke2019pytorch} complied with CUDA (v12.1)(\url{https://pytorch.org}) for model training and evaluation. In pretraining stage, we employed DINOv2 (\url{https://github.com/facebookresearch/dinov2}) for self-supervised pretraining using 2$\times$8 80GB NVIDIA H800 GPU cards. To adapt the size of cell instances, we reduced the local crop scale to 0.02 and set the local crop number to 8. Other pretraining parameters are listed in Extended Data Table. \ref{ST_pretrain_para}. For WSI patching and preprocessing, we employed OpenSlide-python (v1.3.1) and in-house library to load cytology WSIs. Then, the CLAM toolkit (\url{https://github.com/mahmoodlab/CLAM}) was utilized for tiling WSI into 1,200$\times$1,200 patches. 
In training stage, we utilized 8$\times$ 24GB NVIDIA GeForce RTX 3090 GPUs to conduct experiments. For detector development, we implement and compare models under the MMDetection library (v3.0.0) (\url{https://github.com/open-mmlab/mmdetection}). To accelerate cell prediction inference within WSI, we employed distributed data parallelism techniques. The parameters of the employed detector and classifier are detailed in Extended Data Table. \ref{ST_det_para}.

\subsection*{Evaluation metrics and setting}\label{subsec4-8}
We conducted large-scale evaluation of cell-level and WSI-level tasks for cancer screening. In cell-level evaluation, we reported overall performance using metrics AP50 (average precision under 0.50 IoU threshold), mAP (mean average precision under IoU from 0.5 to 0.95, step 0.05), and mAR (mean average recall under IoU from 0.5 to 0.95, step 0.05). We also described the AP50 for fine-grained per-class evaluations.

In WSI-level evaluation, we reported the performance of subgroups including ECA, ASC-US+, LSIL+, HSIL+. The cytological screening results from different subgroups guide subsequent diagnostic procedures and follow-up actions. For instance, if the screening result falls under the ECA subgroup—includes any category of squamous or glandular abnormalities—the next step is to perform an HPV test. If the result falls under the LSIL+ subgroup (indicating LSIL or more severe findings), a colposcopy is recommended to further evaluate the cervical tissue. Therefore, we summed the predicted logits of each class in the subgroup to obtain binary predictions. In retrospective and prospective studies, we adopted metrics consisting of ACC (accuracy), AUC (Area Under ROC Curve), and F1 Score to evaluate the overall performance of WSI classification, and provided sensitivity and specificity metrics. In the experiments, we tuned heyeparameters to ensure gradual convergence during pretraining and finetuning (Extended Data Table. \ref{ST_pretrain_para}-\ref{ST_det_para}). The checkpoint achieving the best performance on the validation set was used for subsequent evaluation experiments.

\subsection*{Statistical analysis}\label{subsec4-11}
We report a 0.95 confidence interval (CI) for statistical analysis. The feature visualization is implemented using t-SNE in scikit-learn (v1.0.2). 

%% file: sec/sec5.tex
\section*{Data availability}\label{sec5-1}
This study involved public and private cytology datasets for system development. The public datasets comprising HErlev (\url{https://mde-lab.aegean.gr/index.php/downloads/}) and SIPaKMeD (\url{https://www.cs.uoi.gr/~marina/sipakmed.html}) are utilized for downstream task evaluation.
The details of private CCS-127K dataset are shown in Fig. \ref{F2_data}(a) and Extended Data Table. \ref{ST_patches}. 

\section*{Code availability}\label{sec5-1}
Source code for model development and the implementation details
will be publicly available upon
paper acceptance. 

\section*{Acknowledgements}\label{sec5-1}
This work was supported by the National Natural Science Foundation of China (No. 62202403), Hong Kong Innovation and Technology Commission (Project No. PRP/034/22FX and ITCPD/17-9), Shenzhen Science and Technology Program (Project No. KCXFZ20230731094059008) and the Project of Hetao Shenzhen-Hong Kong Science and Technology Innovation Cooperation Zone (HZQB-KCZYB-2020083). 

\section*{Author contributions}\label{sec5-1}
H.C., H.J., C.J., and H.L. investigated and designed the study. H.L. and L.D. collected a large-scale clinical cytology dataset. H.L., J.H., and L.D. provided the cell and WSI annotations. Data quality control and preprocessing were completed by H.S., C.J., and Y.Q.. C.J. developed the library for cytology slide reading. Drafting of the original manuscript was contributed by H.J.. C.J. designed and implemented the SSL algorithm. H.J., C.J., and J.M. were responsible for coding and implementing the cell-level and WSI-level screening algorithms. H.J. and Y.Z. developed the adaptation algorithm. H.J. conducted retrospective validation experiments. Clinical guidance was provided by R.C.K.C., A.H., and C.L., who also screened and interpreted WSIs in prospective validation. L.L., X.W., Q.W. and H.C. reviewed and edited the manuscript. Visualization and analysis of experiments were provided by R.L. and Z.C.. H.J. and C.J. contributed to data statistics and analysis. H.C. supervised this research.

\section*{Competing interests}\label{sec5-1}
The authors have no conflicts of
interest to declare.

\section*{Ethics approval}\label{sec5-1}
This study has been reviewed and approved by the Human and Artefacts Research Ethics Committee (HAREC) at the Hong Kong University of Science and Technology, with the protocol number HREP-2024-0216.

\section*{Additional information}\label{sec5-1}

\textbf{Correspondence and requests for materials} should be addressed to Hao Chen.

%% file: sec/sec6.tex
%%%%%%%%%%%%%%%%%%%%%%%%%%%%%%%%%%%%%%%%%%%%%%%%%%%%%%%%%%%%%%%%%%%%%%%%%%%%%%%%%%%%%%%%
\newpage
\section*{Extended Data}\label{secA1}

\setcounter{figure}{0}
\begin{figure*}[h]
\renewcommand{\figurename}{Extended Data Fig.}
    \centering
    \includegraphics[width=1.0\linewidth]{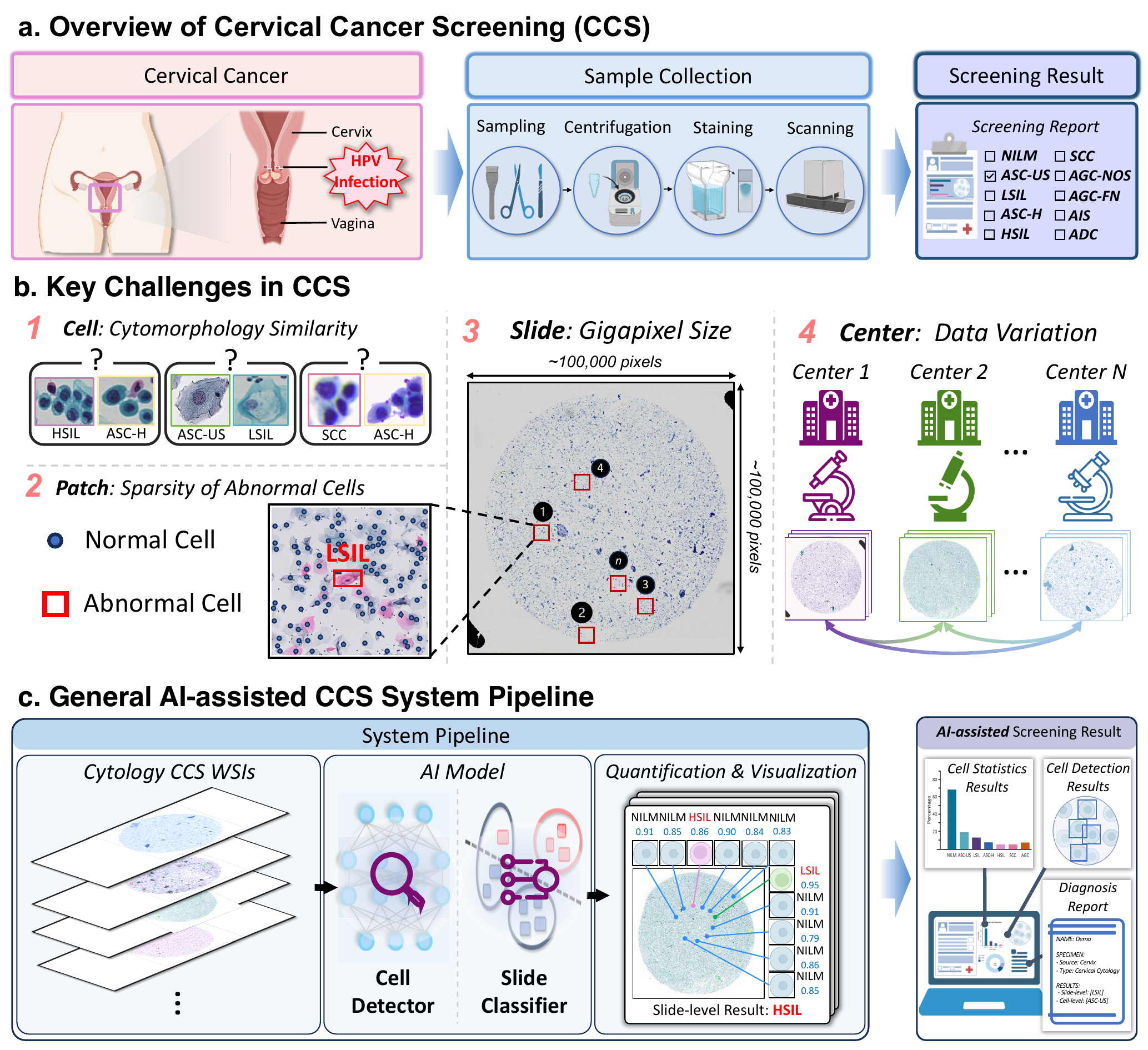}
    \caption{\textbf{Overview of cervical cancer screening and computational cytology.} \textbf{a}. Illustration of cervical cells infected with human papillomavirus (HPV), leading to cervical cancer. The cytology sample collection involves sampling, centrifugation, staining, imaging, for cytologist examinations with screening reports. \textbf{b}. Key challenges in CCS include cytomorphology similarity, sparse abnormal cell distribution, identifying abnormal cells in gigapixel-sized whole slide images (WSI), and data variability. \textbf{c}. A general AI-assisted cancer screening pipeline, comprising a cell detector and a slide classifier, provides quantitative and visualized predictions for both cell-level and slide-level screening.}
    \label{F1_intro}
\end{figure*}

\begin{figure*}[h]
\renewcommand{\figurename}{Extended Data Fig.}
    \centering
    \includegraphics[width=1.0\linewidth]{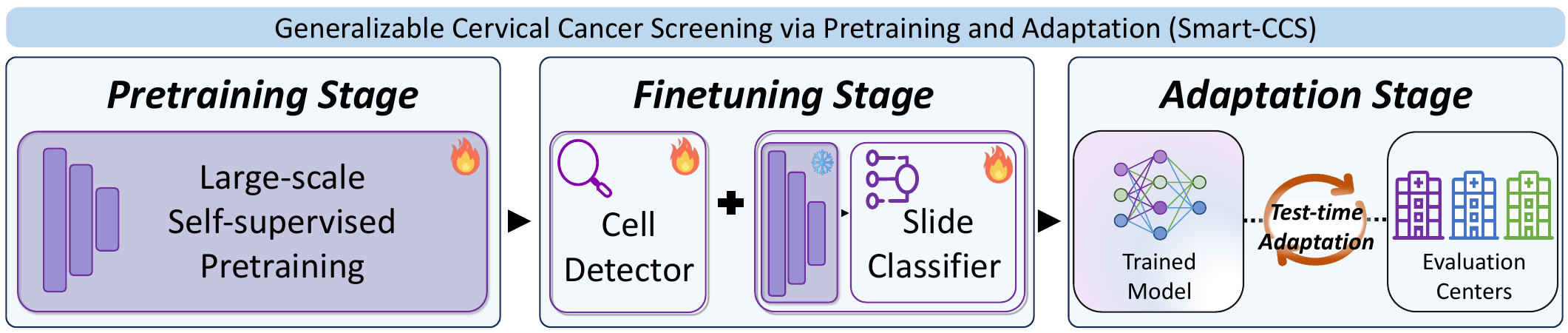}
    \caption{\textbf{The conceptual illustration of proposed Smart-CCS paradigm.} It consists of three sequential stages: 1) large-scale self-supervised pretraining, 2) CCS model finetuning, and 3) test-time adaptation. }
    \label{SF_method}
\end{figure*}

\clearpage
\begin{figure*}[h]
\renewcommand{\figurename}{Extended Data Fig.}
    \centering
    \includegraphics[width=1.0\linewidth]{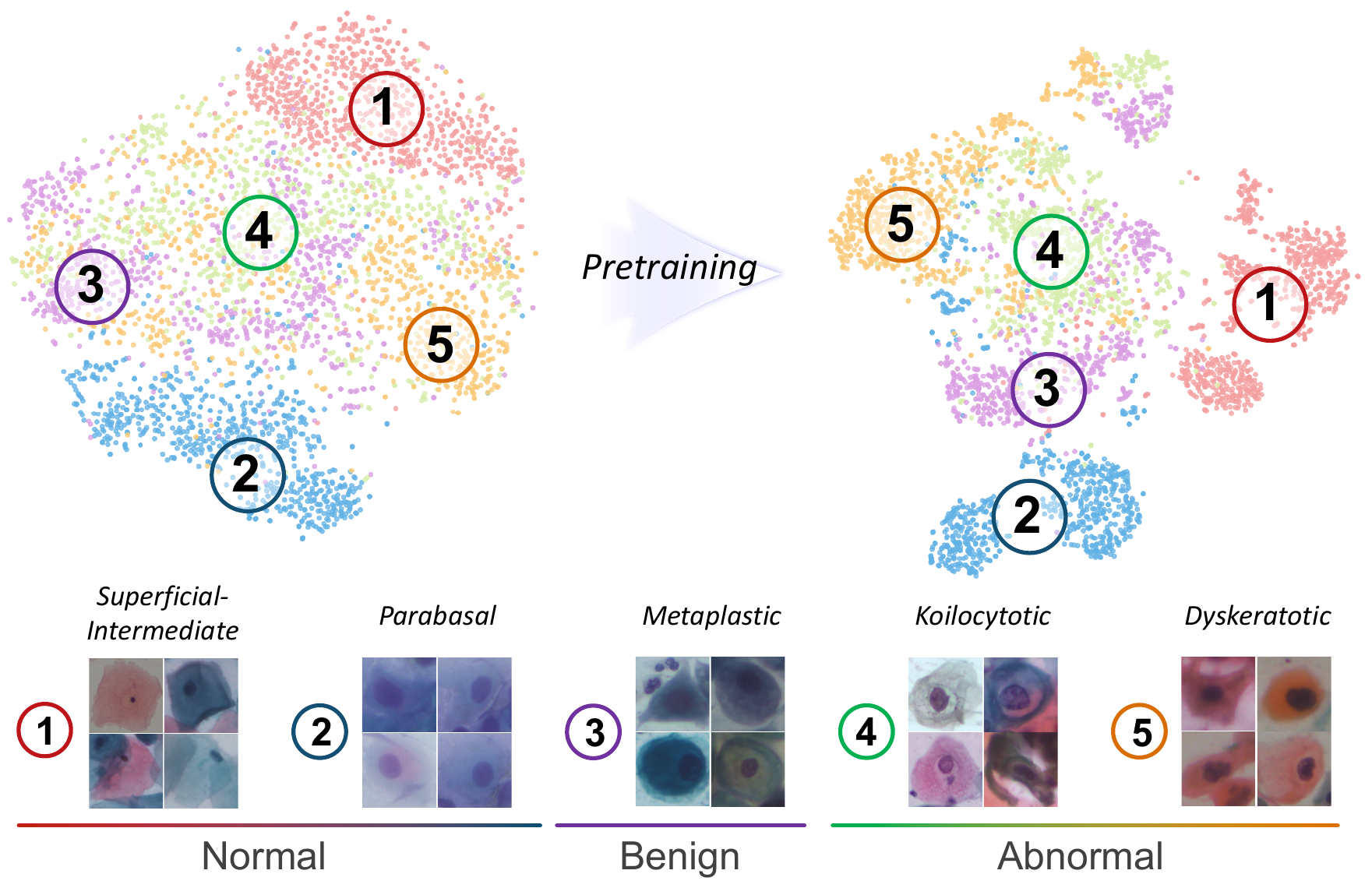}
    \caption{\textbf{Visualization of the effectiveness of pretraining from SIPaKMeD \cite{plissiti2018sipakmed} data using t-SNE, showcasing cervical cell images and feature distributions both without and with pretraining.} Five colors (dark blue, light blue, light orange, deep orange, green) denote five cell categories: superficial-intermediate (normal), parabasal (normal), metaplastic (benign), koilocytotic (abnormal), dyskeratotic (abnormal). Cell images are extracted with LVD-142M pretrained ViT-Large and our CCS-127K pretrained ViT-Large encoder to obtain 1024-dimensional features, subsequently reduced by t-SNE and plotted as a scatter plot. }
    \label{SF_tsne}
\end{figure*}

\clearpage
\begin{figure*}[h]
\renewcommand{\figurename}{Extended Data Fig.}
    \centering
    \includegraphics[width=1.0\linewidth]{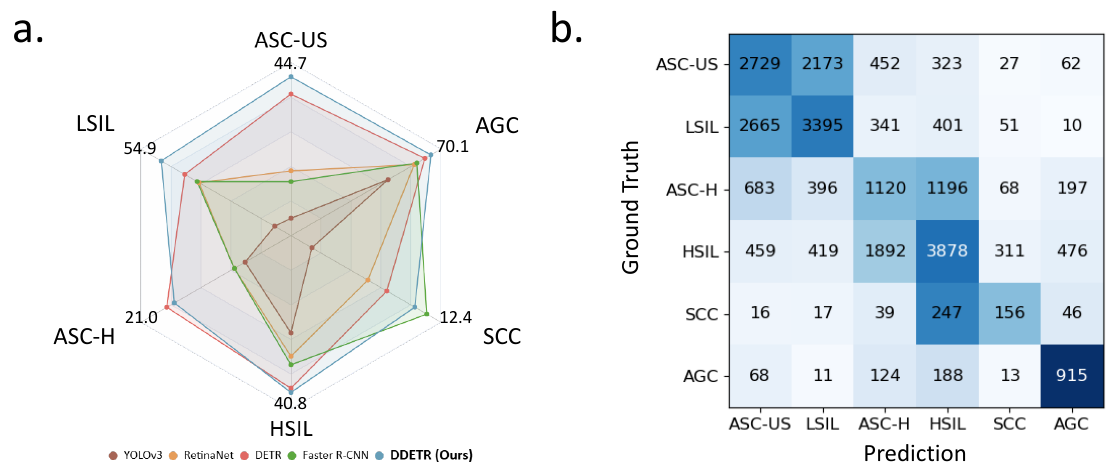}
    \caption{\textbf{Cervical cell detection and classification performance.} a) Performance comparison of per-class AP50 of detectors, demonstrating consistent performance improvement across all categories. b) Confusion matrix illustrating the classification performance of detectors, showing distinguishability between cell types such as glandular cells and squamous cells, as well as category correlations between ASC-US and LSIL, ASC-H and HSIL.}
    \label{SF_det}
\end{figure*}

\clearpage
\begin{figure*}[h]
\renewcommand{\figurename}{Extended Data Fig.}
    \centering
    \includegraphics[width=1.0\linewidth]{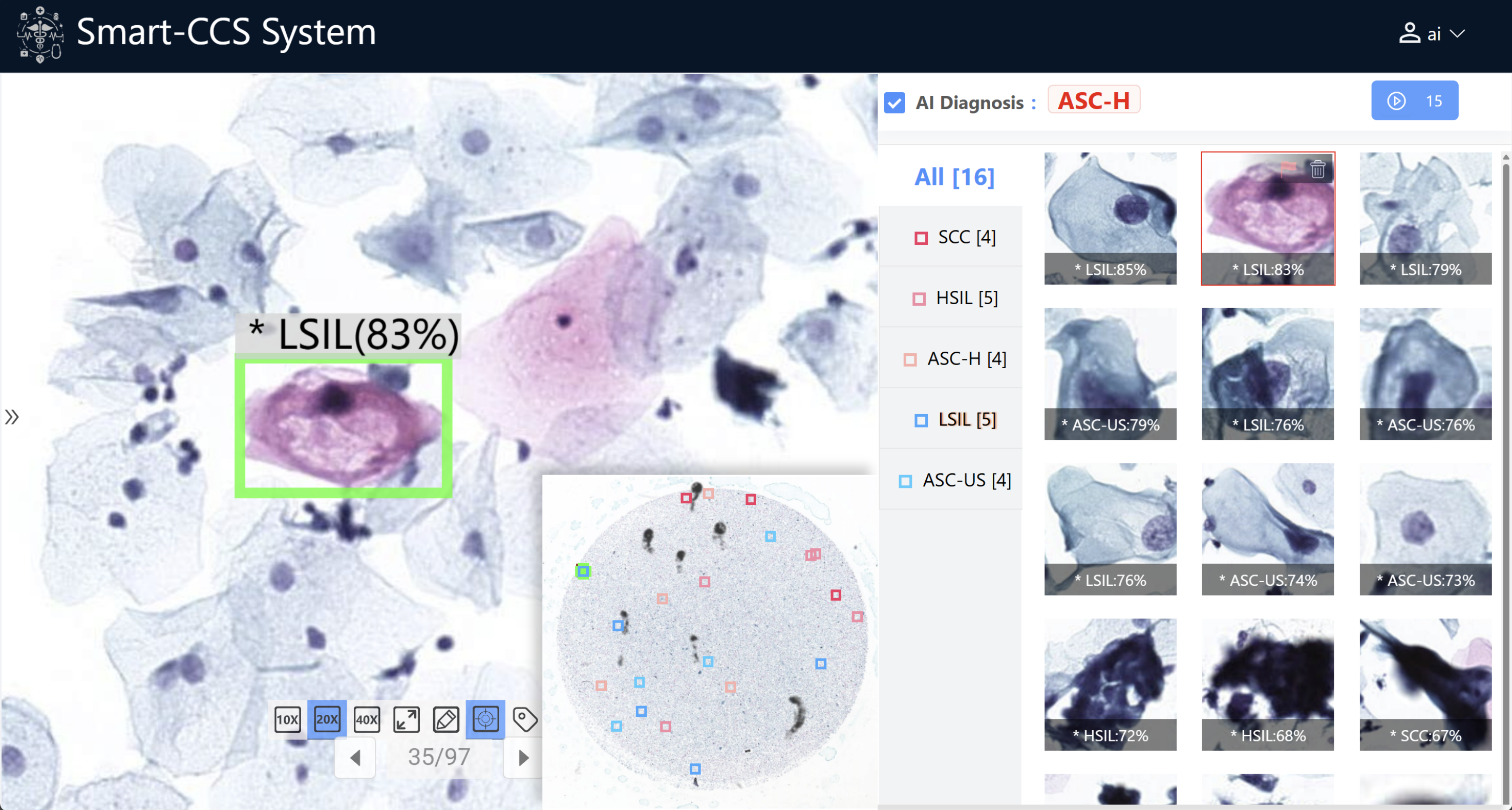}
    \caption{\textbf{The integrated and interactive interface of the developed cervical cancer screening system.} This integrated CCS system provides both local and global views of cytology samples, highlighting screened suspicious cells (LSIL, HSIL, etc.) and ultimately delivering TBS diagnostic suggestions (ASC-H).
    }
    \label{SF_integrate}
\end{figure*}

\clearpage
\begin{figure*}[h]
\renewcommand{\figurename}{Extended Data Fig.}
    \centering
    \includegraphics[width=1.0\linewidth]{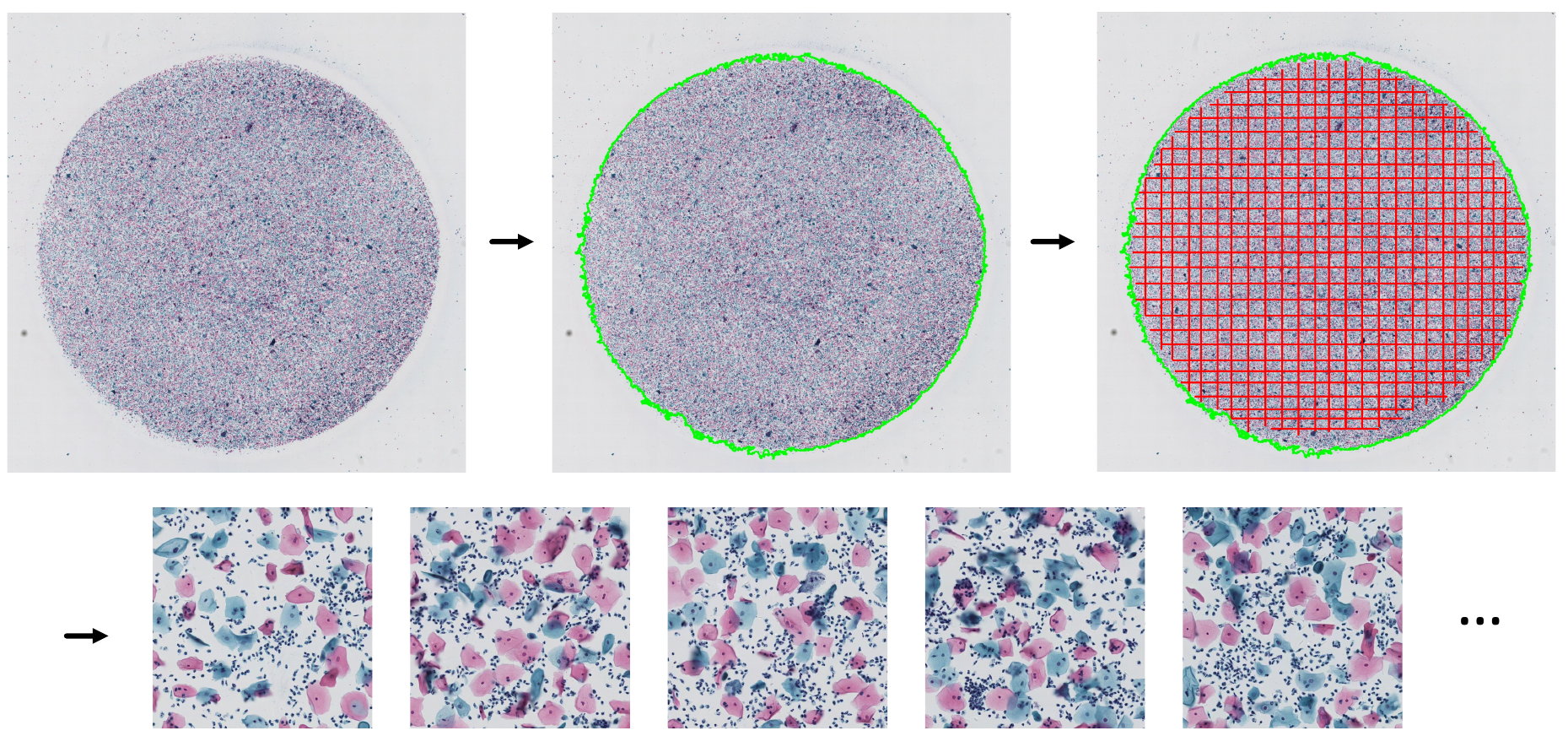}
    \caption{\textbf{Whole slide image preprocessing.} Central circular foreground is extracted from the whole-slide image (69,888 pixels in the given example) using CLAM toolkit \cite{lu2021data}, and then cut into 1,200$\times$1,200 patches for further processing. }
    \label{SF_preprocessing}
\end{figure*}

\clearpage
\begin{figure*}[h]
\renewcommand{\figurename}{Extended Data Fig.}
    \centering
    \includegraphics[width=1.0\linewidth]{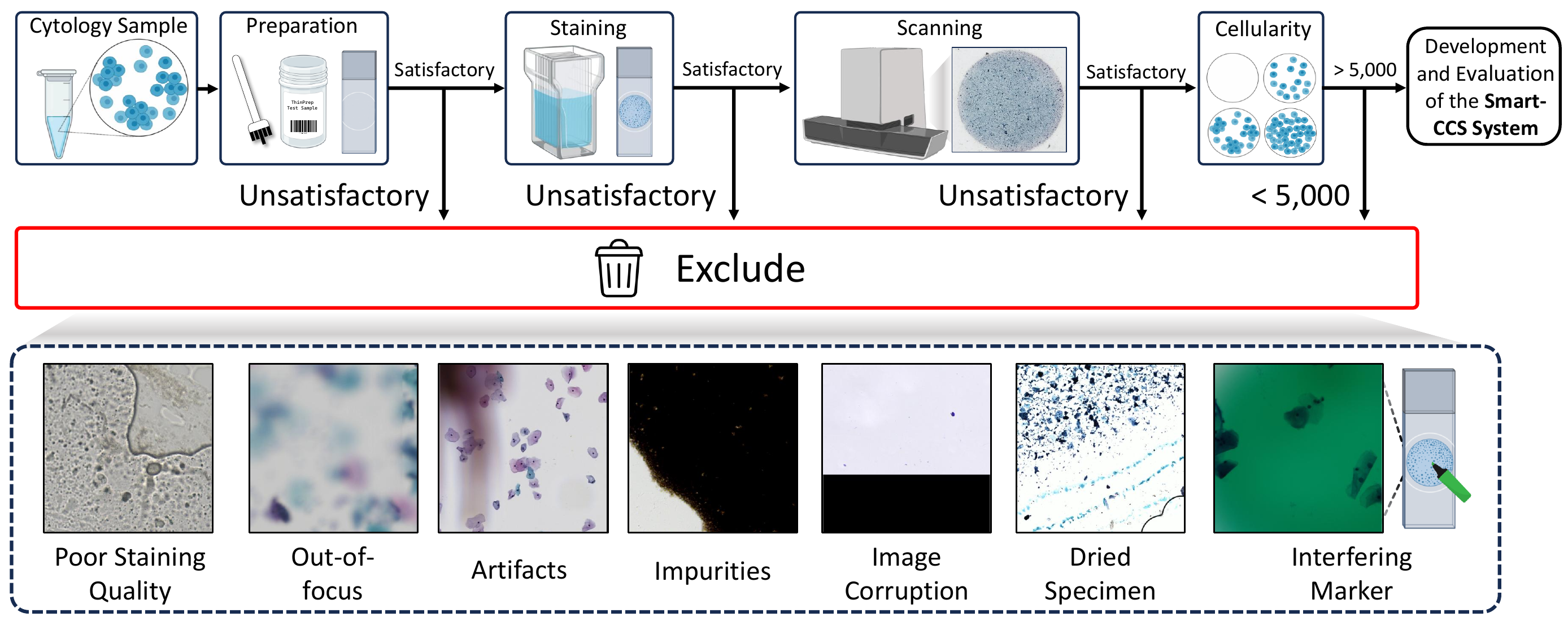}
    \caption{\textbf{Workflow of quality control for WSI.} Quality control for WSI involves considerations of cellularity ($>$5,000 per slide \cite{nayar2015bethesda}), preparation, staining, and scanning. Typical samples that are excluded include poor staining quality, out-of-focus, artifacts, impurities, image corruption, dried specimens, and interfering labeling.}
    \label{SF_qualitycontrol}
\end{figure*}

\clearpage
\begin{table}[h] 
\renewcommand{\arraystretch}{1.5}
\renewcommand{\tablename}{Extended Data Table.}
\centering 
\caption{\textbf{Abbreviations and nomenclature in this study.}}
\begin{tabular}{c|c} 
\hline
\rowcolor{cusyellow} \textbf{Abbreviations} & \textbf{Nomenclature} \\ 
\hline
 CC & Cervical cancer \\ 
\rowcolor{cusyellowl} CCS & Cervical cancer screening \\ 
 NILM & Negative for intraepithelial lesion or malignancy \\ 
\rowcolor{cusyellowl} ASC-US & Atypical squamous cells of undetermined significance \\ 
 LSIL & Low–grade squamous intraepithelial lesion \\ 
\rowcolor{cusyellowl} ASC-H & Atypical squamous cells cannot exclude an HSIL \\ 
 HSIL & High–grade squamous intraepithelial lesions \\ 
\rowcolor{cusyellowl} SCC & Squamous cell carcinoma \\ 
 AGC & Atypical glandular cells \\ 
\rowcolor{cusyellowl} AGC-FN & Atypical glandular cells, favor neoplastic \\ 
 AGC-NOS & Atypical glandular cells, not otherwise specified \\ 
\rowcolor{cusyellowl} AIS & Adenocarcinoma in situ \\ 
 ADC & Adenocarcinoma \\ 
\rowcolor{cusyellowl} TBS & The Bethesda system \\ 
 SIL & Squamous intraepithelial lesion \\ 
\rowcolor{cusyellowl} HPV & Human papillomavirus \\ 
 ECA & Epithelial cell abnormalities \\ 
\rowcolor{cusyellowl} CIN & Cervical intraepithelial neoplasia \\ 
\hline 
\end{tabular} 
\label{ST_abb}
\end{table}

\clearpage
\begin{table}[h] 
\renewcommand{\arraystretch}{1.5}
\renewcommand{\tablename}{Extended Data Table.}
\centering 
\caption{\textbf{Investigation of the effectiveness and scaling laws of self-supervised pretraining for cell classification tasks (Top-1 accuracy)}. Three cell classification datasets are evaluated, SIPaKMeD\cite{plissiti2018sipakmed}, HErlev \cite{jantzen2005pap}, and CCS-Cell. The best results are highlighted in bold.}
\begin{tabular}{c|ccc} 
\hline
\rowcolor{cusyellow} \textbf{Pretrain settings }& \textbf{SIPaKMeD} &\textbf{HErlev} & \textbf{CCS-Cell} \\ 
\hline
0M (w/o Pretrain)&0.926 (0.908 - 0.944) & 0.842 (0.789 - 0.895)&0.827 (0.810 - 0.844)\\
\rowcolor{cusyellowl} 1M& 0.936 (0.919 - 0.953) & 0.887 (0.841 - 0.933) & 0.859 (0.843 - 0.875) \\
6M&0.937 (0.920 - 0.954) & 0.890 (0.845 - 0.935)&0.863 (0.847 - 0.879)\\
\rowcolor{cusyellowl} 30M&0.940 (0.924 - 0.956) & 0.895 (0.851 - 0.939)&0.872 (0.857 - 0.887)\\
60M&0.942 (0.926 - 0.958) & 0.905 (0.863 - 0.947)&\textbf{0.883 (0.868 - 0.898)}\\
\rowcolor{cusyellowl} 100M&\textbf{0.960 (0.946 - 0.974)} & \textbf{0.914 (0.873 - 0.955)}&\textbf{0.883 (0.868 - 0.898)}\\
\hline
\end{tabular} 
\label{ST_pretrain_cell}
\end{table}

\clearpage
\begin{table}[h] 
\renewcommand{\arraystretch}{1.5}
\renewcommand{\tablename}{Extended Data Table.}
\centering 
\caption{\textbf{Investigation of the effectiveness and scaling laws of self-supervised pretraining for WSI classification tasks (Top-1 accuracy)}. Three retrospective centers are included, and we report the overall fine-grained classification results (ALL) along with the performance of different groups (ECA, ASC-US+, LSIL+, and HSIL+).}
\begin{tabular}{cc|ccc} 
\hline
\rowcolor{cusyellow} \multicolumn{2}{c|}{\multirow{1}{*}{\textbf{Pretrain settings}}} & \textbf{CCS-127K RC2}& \textbf{CCS-127K RC4} & \textbf{CCS-127K RC5} \\ 
\hline
\multirow{1}{*}{0M }&ECA &0.945 (0.936 - 0.954) &0.857 (0.840 - 0.873)& 0.712 (0.684 - 0.740)\\
& ASC-US+  & 0.940 (0.930 - 0.949) & 0.841 (0.824 - 0.858) & 0.692 (0.664 - 0.721)\\
& LSIL+ & 0.949 (0.940 - 0.958) & 0.842 (0.824 - 0.859) & 0.788 (0.763 - 0.813) \\
& HSIL+ & 0.986 (0.981 - 0.991) & 0.932 (0.920 - 0.944) & 0.887 (0.868 - 0.906) \\
& ALL & 0.895 (0.883 - 0.908) & 0.687 (0.665 - 0.709) & 0.638 (0.608 - 0.667) \\

 \rowcolor{cusyellowl} \multirow{1}{*}{1M}&ECA & 0.943 (0.934 - 0.952) & 0.887 (0.872 - 0.899) & 0.932 (0.916 - 0.947) \\
\rowcolor{cusyellowl} & ASC-US+ & 0.935 (0.925 - 0.944) & 0.873 (0.857 - 0.886) & 0.917 (0.900 - 0.934) \\
\rowcolor{cusyellowl} & LSIL+  & 0.942 (0.932 - 0.951) & 0.837 (0.820 - 0.852)  & 0.876 (0.856 - 0.896) \\
\rowcolor{cusyellowl} & HSIL+   & 0.983 (0.978 - 0.988) & 0.916 (0.903 - 0.927)  & 0.946 (0.933 - 0.960)\\
\rowcolor{cusyellowl} & ALL     & 0.889 (0.876 - 0.902) & 0.711 (0.689 - 0.729)  & 0.792 (0.767 - 0.816)\\

\multirow{1}{*}{6M}&ECA  & 0.952 (0.943 - 0.961) & 0.875 (0.859 - 0.891)& 0.937 (0.922 - 0.952) \\
&ASC-US+  & 0.942 (0.933 - 0.951) & 0.858 (0.842 - 0.875) & 0.927 (0.911 - 0.943)\\
&LSIL+    & 0.945 (0.935 - 0.954) & 0.832 (0.814 - 0.850) & 0.895 (0.876 - 0.914)\\
&HSIL+   & 0.985 (0.980 - 0.990) & 0.917 (0.904 - 0.930) & 0.948 (0.935 - 0.962) \\
&ALL     & 0.897 (0.884 - 0.909) & 0.693 (0.671 - 0.715) & 0.808 (0.784 - 0.832) \\

  \rowcolor{cusyellowl} \multirow{1}{*}{30M}&ECA & 0.952 (0.944 - 0.961) & 0.882 (0.866 - 0.897) & 0.944 (0.929 - 0.958) \\
  \rowcolor{cusyellowl} &ASC-US+ & 0.945 (0.936 - 0.954) & 0.868 (0.852 - 0.884) & 0.936 (0.921 - 0.951) \\
  \rowcolor{cusyellowl} &LSIL+  & 0.945 (0.936 - 0.954) & 0.848 (0.830 - 0.865)  & 0.884 (0.865 - 0.904) \\
  \rowcolor{cusyellowl} &HSIL+    & 0.984 (0.979 - 0.989) & 0.927 (0.915 - 0.940) & 0.946 (0.933 - 0.960)\\
  \rowcolor{cusyellowl} &ALL    & 0.896 (0.884 - 0.908) & 0.717 (0.695 - 0.738) & 0.808 (0.784 - 0.832) \\ 

\multirow{1}{*}{60M}&ECA   & 0.954 (0.946 - 0.963) & 0.915 (0.902 - 0.929)& 0.949 (0.936 - 0.963)\\
&ASC-US+  & 0.947 (0.938 - 0.956) & 0.900 (0.886 - 0.914)& 0.941 (0.926 - 0.955)\\
&LSIL+   & 0.947 (0.938 - 0.956) & 0.857 (0.840 - 0.874)  &  0.905 (0.887 - 0.923)\\
&HSIL+  & 0.985 (0.981 - 0.990) & 0.930 (0.918 - 0.942)  & 0.946 (0.932 - 0.959) \\
&ALL & 0.907 (0.895 - 0.919) & 0.742 (0.721 - 0.763) & 0.816 (0.792 - 0.840) \\

  \rowcolor{cusyellowl} \multirow{1}{*}{100M}&ECA  & 0.950 (0.942 - 0.959) & 0.915 (0.902 - 0.929) & 0.958 (0.946 - 0.970) \\
  \rowcolor{cusyellowl} &ASC-US+  & 0.942 (0.933 - 0.951) & 0.898 (0.884 - 0.913) & 0.949 (0.936 - 0.963) \\
  \rowcolor{cusyellowl} &LSIL+ & 0.951 (0.943 - 0.960) & 0.855 (0.838 - 0.872) & 0.914 (0.897 - 0.931)\\
  \rowcolor{cusyellowl} &HSIL+   & 0.985 (0.981 - 0.990) & 0.927 (0.915 - 0.940) & 0.956 (0.944 - 0.969)\\
  \rowcolor{cusyellowl} &ALL & 0.902 (0.890 - 0.913) & 0.742 (0.721 - 0.763) & 0.835 (0.812 - 0.857) \\ 

\hline 
\end{tabular} 
\label{ST_pretrain_wsi}
\end{table}

\clearpage
\begin{table}[h] 
\renewcommand{\arraystretch}{1.5}
\renewcommand{\tablename}{Extended Data Table.}
\centering 
\caption{\textbf{Ablation experiments for self-supervised pretraining backbone (ViT-Large and ViT-Gaint) and algorithm (DINOv2 \cite{oquabdinov2} and MoCov3 \cite{chen2021empirical}) (Top-1 accuracy)}. The best results are in bold.}
\begin{tabular}{c|ccc} 
\hline
 \rowcolor{cusyellow} \textbf{Pretrain settings} & \textbf{SIPaKMeD }&\textbf{HErlev} & \textbf{CCS-Cell} \\ 
\hline
w/o Pretrain&0.926 (0.908 - 0.944) & 0.842 (0.789 - 0.895)&0.827 (0.810 - 0.844)\\
 \rowcolor{cusyellowl} ViT-Large+DINOv2 &\textbf{0.942 (0.926 - 0.958)} & 0.905 (0.863 - 0.947)&\textbf{0.883 (0.868 - 0.898)}\\
ViT-Gaint+DINOv2 &0.934 (0.917 - 0.951) & \textbf{0.906 (0.864 - 0.948)}&0.877 (0.862 - 0.892)\\
 \rowcolor{cusyellowl}ViT-Large+MoCov3 &0.927 (0.909 - 0.945) & 0.892 (0.847 - 0.937)&0.868 (0.852 - 0.884)\\
\hline 
\end{tabular} 
\label{ST_pretrain_ablation}
\end{table}

\clearpage
\begin{table}[h] 
\renewcommand{\arraystretch}{1.5}
\renewcommand{\tablename}{Extended Data Table.}
\centering 
\caption{\textbf{Overall performance and comparison of SOTA  methods for abnormal cell detection.} The best results are in bold.}
    \begin{tabular}{c|ccc}
    \hline
    \rowcolor{cusyellow} \textbf{Model} & \textbf{mAP} & \textbf{AP50} & \textbf{mAR} \\ 
    \hline
    \textbf{YOLOv3} & 0.134 (0.122–0.146) & 0.263 (0.248–0.278) & 0.349 (0.333–0.365) \\ 
     \rowcolor{cusyellowl} \textbf{Faster R-CNN} & 0.206 (0.192–0.220) & 0.347 (0.331–0.363) & 0.432 (0.415–0.449) \\ 
    \textbf{RetinaNet} & 0.201 (0.187–0.215) & 0.333 (0.317–0.349) & 0.463 (0.446–0.480) \\ 
     \rowcolor{cusyellowl} \textbf{DETR} & 0.209 (0.195–0.223) & 0.390 (0.374–0.406) & 0.470 (0.453–0.487) \\ 
    \textbf{DDETR} & \textbf{0.239 (0.225–0.253)} & \textbf{0.406 (0.389–0.423)} & \textbf{0.490 (0.473–0.507)} \\ 
    \hline
\end{tabular}
\label{ST_det}
\end{table}

\begin{table}[h] 
\renewcommand{\arraystretch}{1.5}
\renewcommand{\tablename}{Extended Data Table.}
\centering 
\caption{\textbf{Class-wise performance and comparison of SOTA methods for abnormal cell detection.} The best results are in bold.}
    \begin{tabular}{c|ccc}
    \hline
    \rowcolor{cusyellow} \textbf{Model} & \textbf{ASC-US} & \textbf{LSIL} & \textbf{ASC-H} \\ 
    \hline
    \textbf{YOLOv3} & 0.316 (0.300–0.332) & 0.375 (0.359–0.391) & 0.125 (0.114–0.136) \\ 
    \rowcolor{cusyellowl} \textbf{Faster R-CNN} & 0.350 (0.334–0.366) & 0.494 (0.477–0.511) & 0.138 (0.126–0.150) \\ 
    \textbf{RetinaNet} & 0.360 (0.344–0.376) & 0.491 (0.474–0.508) & 0.138 (0.126–0.150) \\ 
    \rowcolor{cusyellowl} \textbf{DETR} & 0.431 (0.414–0.448) & 0.513 (0.496–0.530) & \textbf{0.219 (0.205–0.233)} \\ 
    \textbf{DDETR} & \textbf{0.447 (0.430–0.464)} & \textbf{0.549 (0.532–0.566)} & 0.210 (0.196–0.224) \\ 
    \hline
\end{tabular}
\label{ST_det_p1}
\end{table}

\begin{table}[h] 
\renewcommand{\arraystretch}{1.5}
\renewcommand{\tablename}{Extended Data Table.}
\centering 
\caption{\textbf{Class-wise performance and comparison of SOTA methods for abnormal cell detection (continued).} The best results are in bold.}
    \begin{tabular}{c|ccc}
    \hline
    \rowcolor{cusyellow} \textbf{Model} & \textbf{HSIL} & \textbf{SCC} & \textbf{AGC} \\ 
    \hline
    \textbf{YOLOv3} & 0.253 (0.238–0.268) & 0.021 (0.016–0.026) & 0.488 (0.471–0.505) \\ 
    \rowcolor{cusyellowl} \textbf{Faster R-CNN} & 0.336 (0.320–0.352) & 0.136 (0.124–0.148) & 0.631 (0.615–0.647) \\ 
    \textbf{RetinaNet} & 0.314 (0.298–0.330) & 0.077 (0.068–0.086) & 0.618 (0.602–0.634) \\ 
    \rowcolor{cusyellowl} \textbf{DETR} & 0.397 (0.380–0.414) & 0.096 (0.086–0.106) & 0.671 (0.655–0.687) \\ 
    \textbf{DDETR} & \textbf{0.408 (0.391–0.425)} & \textbf{0.124 (0.113–0.135)} & \textbf{0.701 (0.686–0.716)} \\ 
    \hline
\end{tabular}
\label{ST_det_p2}
\end{table}

\clearpage
\begin{table}[h] 
\renewcommand{\arraystretch}{2}
\renewcommand{\tablename}{Extended Data Table.}
\centering 
\caption{\textbf{Internal testing results of Smart-CCS in retrospective study (RC1-RC3).}}
\begin{tabular}{cc|c|c|c} 
\hline
\rowcolor{cusyellow} \multicolumn{2}{c|}{\multirow{1}{*}{\textbf{
 Metrics}}} & \textbf{RC1} & \textbf{RC2} & \textbf{RC3} \\ 
\hline
\multirow{1}{*}{ECA}&Accuracy (95\% CI) & 0.663 (0.644–0.682) & 0.940 (0.931–0.949) & 0.968 (0.960–0.976) \\
& AUC (95\% CI) & 0.767 (0.750–0.784) & 0.971 (0.965–0.978) & 0.990 (0.985–0.995) \\
& F1 Score (95\% CI) & 0.710 (0.692–0.728) & 0.940 (0.931–0.950) & 0.968 (0.960–0.976) \\
& Sensitivity (95\% CI) & 0.713 (0.695–0.731) & 0.859 (0.845–0.873) & 0.938 (0.927–0.949) \\
& Specificity (95\% CI) & 0.654 (0.635–0.673) & 0.961 (0.954–0.969) & 0.978 (0.971–0.985) \\

 \rowcolor{cusyellowl}  \multirow{1}{*}{ASC-US+ }&Accuracy (95\% CI) & 0.726 (0.709–0.743) & 0.934 (0.924–0.944) & 0.967 (0.959–0.975) \\
 \rowcolor{cusyellowl}  & AUC (95\% CI) & 0.767 (0.750–0.784) & 0.967 (0.960–0.974) & 0.990 (0.985–0.995)\\
 \rowcolor{cusyellowl}  & F1 Score (95\% CI)  & 0.759 (0.742–0.776) & 0.934 (0.925–0.944) & 0.967 (0.959–0.975)  \\
 \rowcolor{cusyellowl}  & Sensitivity (95\% CI) & 0.626 (0.607–0.645) & 0.852 (0.838–0.866) & 0.932 (0.920–0.944) \\
 \rowcolor{cusyellowl}  & Specificity (95\% CI)      & 0.743 (0.726–0.760) & 0.954 (0.946–0.962) & 0.978 (0.971–0.985) \\

\multirow{1}{*}{LSIL+}&Accuracy (95\% CI) & 0.807 (0.792–0.822) & 0.911 (0.900–0.922) & 0.848 (0.831–0.865) \\
& AUC (95\% CI) & 0.902 (0.890–0.914) & 0.975 (0.969–0.982) & 0.926 (0.914–0.938) \\
& F1 Score (95\% CI) & 0.859 (0.845–0.873) & 0.923 (0.913–0.933) & 0.896 (0.882–0.910)  \\
& Sensitivity (95\% CI) & 0.816 (0.801–0.831) & 0.929 (0.919–0.939) & 0.957 (0.948–0.966)  \\
& Specificity (95\% CI) & 0.807 (0.792–0.822) & 0.909 (0.898–0.921) & 0.844 (0.827–0.861) \\

 \rowcolor{cusyellowl}   \multirow{1}{*}{HSIL+}&Accuracy (95\% CI) & 0.967 (0.960–0.974) & 0.970 (0.964–0.977) & 0.898 (0.884–0.912)  \\
 \rowcolor{cusyellowl}  & AUC (95\% CI) & 0.943 (0.934–0.952) & 0.982 (0.977–0.987) & 0.934 (0.923–0.945) \\
 \rowcolor{cusyellowl}  & F1 Score (95\% CI)  & 0.979 (0.973–0.985) & 0.976 (0.970–0.982) & 0.936 (0.925–0.947) \\
 \rowcolor{cusyellowl}  & Sensitivity (95\% CI) & 0.750 (0.733–0.767) & 0.826 (0.811–0.841) & 0.800 (0.782–0.818)\\
 \rowcolor{cusyellowl}  & Specificity (95\% CI) & 0.968 (0.961–0.975) & 0.973 (0.967–0.980) & 0.899 (0.885–0.913)\\
\hline
\end{tabular} 
\label{ST_wsi}
\end{table}

\clearpage
\begin{table}[h] 
\renewcommand{\arraystretch}{2}
\renewcommand{\tablename}{Extended Data Table.}
\centering 
\caption{\textbf{Internal testing results of Smart-CCS in retrospective study (RC4-RC6).}}
\begin{tabular}{cc|c|c|c} 
\hline
 \rowcolor{cusyellow} \multicolumn{2}{c|}{\multirow{1}{*}{\textbf{
 Metrics}}} & \textbf{RC4} & \textbf{RC5} & \textbf{RC6}\\ 
\hline
\multirow{1}{*}{ECA}&Accuracy (95\% CI) & 0.894 (0.879–0.908) & 0.943 (0.928–0.957) & 0.821 (0.796–0.846)  \\
& AUC (95\% CI) & 0.968 (0.959–0.976) & 0.985 (0.977–0.992) & 0.890 (0.870–0.910) \\
& F1 Score (95\% CI) & 0.894 (0.879–0.908) & 0.943 (0.928–0.957) & 0.839 (0.816–0.863)  \\
& Sensitivity (95\% CI) & 0.961 (0.951–0.970) & 0.948 (0.934–0.961) & 0.814 (0.789–0.839) \\
& Specificity (95\% CI) & 0.837 (0.819–0.855) & 0.938 (0.923–0.952) & 0.822 (0.798–0.847)  \\

 \rowcolor{cusyellowl}  \multirow{1}{*}{ASC-US+ }&Accuracy (95\% CI) & 0.877 (0.861–0.893) & 0.925 (0.908–0.941) & 0.823 (0.799–0.848) \\
 \rowcolor{cusyellowl}  & AUC (95\% CI)  & 0.957 (0.947–0.967) & 0.978 (0.968–0.987) & 0.888 (0.868–0.909)  \\
 \rowcolor{cusyellowl}  & F1 Score (95\% CI)  & 0.877 (0.862–0.893)& 0.925 (0.908–0.941) & 0.841 (0.817–0.864) \\
 \rowcolor{cusyellowl}  & Sensitivity (95\% CI) & 0.961 (0.952–0.971) & 0.947 (0.933–0.961) & 0.800 (0.774–0.826) \\
 \rowcolor{cusyellowl}  & Specificity (95\% CI)   & 0.812 (0.794–0.831)    & 0.905 (0.887–0.923) & 0.828 (0.803–0.852) ) \\

\multirow{1}{*}{LSIL+}&Accuracy (95\% CI) & 0.828 (0.810–0.846) & 0.870 (0.849–0.891) & 0.831 (0.807–0.855) \\
& AUC (95\% CI) & 0.916 (0.903–0.930) & 0.965 (0.953–0.976) & 0.946 (0.931–0.960) \\
& F1 Score (95\% CI) & 0.838 (0.820–0.855)& 0.875 (0.855–0.896) & 0.864 (0.842–0.886)  \\
& Sensitivity (95\% CI)& 0.842 (0.824–0.859) & 0.957 (0.945–0.970) & 0.935 (0.919–0.951) \\
& Specificity (95\% CI) & 0.824 (0.806–0.842) & 0.837 (0.814–0.860) & 0.821 (0.797–0.846) \\

 \rowcolor{cusyellowl}   \multirow{1}{*}{HSIL+}&Accuracy (95\% CI) & 0.878 (0.862–0.894)& 0.911 (0.893–0.928) & 0.909 (0.891–0.928) \\
 \rowcolor{cusyellowl}  & AUC (95\% CI) & 0.918 (0.904–0.931) & 0.972 (0.962–0.983) & 0.973 (0.963–0.984)  \\
 \rowcolor{cusyellowl}  & F1 Score (95\% CI)   & 0.895 (0.880–0.909)& 0.918 (0.901–0.935) & 0.930 (0.913–0.946) \\
 \rowcolor{cusyellowl}  & Sensitivity (95\% CI)  & 0.818 (0.800–0.837)& 0.895 (0.876–0.914) & 0.897 (0.878–0.917) \\
 \rowcolor{cusyellowl}  & Specificity (95\% CI)  &  0.884 (0.869–0.899) & 0.913 (0.896–0.930) & 0.910 (0.892–0.928) \\

\hline 
\end{tabular} 
\label{ST_wsi_2}
\end{table}

\clearpage
\begin{table}[h] 
\renewcommand{\arraystretch}{2}
\renewcommand{\tablename}{Extended Data Table.}
\centering 
\caption{\textbf{Internal testing results of Smart-CCS in retrospective study (RC7-RC17).} Note: R14-R17 were merged due to the limited sample size at these four centers.}
\begin{tabular}{cc|c|c} 
\hline
 \rowcolor{cusyellow} \multicolumn{2}{c|}{\multirow{1}{*}{\textbf{
 Metrics}}}  & \textbf{RC7 }& \textbf{RC14-17 }\\ 
\hline
\multirow{1}{*}{ECA}&Accuracy (95\% CI) & 0.942 (0.926–0.958) & 0.968 (0.952–0.984) \\
& AUC (95\% CI)  & 0.971 (0.960–0.982) & 0.971 (0.956–0.986) \\
& F1 Score (95\% CI) & 0.942 (0.926–0.958) & 0.968 (0.952–0.984) \\
& Sensitivity (95\% CI)  & 0.857 (0.833–0.881) & 0.940 (0.918–0.961) \\
& Specificity (95\% CI)  & 0.968 (0.956–0.980) & 0.977 (0.964–0.991) \\

 \rowcolor{cusyellowl}  \multirow{1}{*}{ASC-US+ }&Accuracy (95\% CI) & 0.948 (0.933–0.963) & 0.970 (0.955–0.986) \\
 \rowcolor{cusyellowl}  & AUC (95\% CI)  & 0.971 (0.960–0.982) & 0.980 (0.967–0.992) \\
 \rowcolor{cusyellowl}  & F1 Score (95\% CI)  & 0.948 (0.933–0.963) & 0.970 (0.955–0.986) \\
 \rowcolor{cusyellowl}  & Sensitivity (95\% CI)  & 0.887 (0.866–0.908) & 0.947 (0.927–0.968) \\
 \rowcolor{cusyellowl}  & Specificity (95\% CI)    & 0.965 (0.953–0.977) & 0.978 (0.964–0.991) \\

\multirow{1}{*}{LSIL+}&Accuracy (95\% CI)  & 0.818 (0.792–0.844) & 0.968 (0.952–0.984) \\
& AUC (95\% CI) & 0.888 (0.867–0.909) & 0.996 (0.989–1.000) \\
& F1 Score (95\% CI) & 0.846 (0.822–0.870) & 0.969 (0.953–0.985) \\
& Sensitivity (95\% CI) & 0.795 (0.768–0.822) & 1.000 (1.000–1.000) \\
& Specificity (95\% CI) & 0.820 (0.794–0.846) & 0.960 (0.942–0.977) \\

 \rowcolor{cusyellowl}   \multirow{1}{*}{HSIL+}&Accuracy (95\% CI) & 0.890 (0.869–0.911) & 0.921 (0.897–0.946) \\
 \rowcolor{cusyellowl}  & AUC (95\% CI)  & 0.932 (0.915–0.949) & 0.994 (0.987–1.001) \\
 \rowcolor{cusyellowl}  & F1 Score (95\% CI)  & 0.913 (0.894–0.932) & 0.939 (0.917–0.961) \\
 \rowcolor{cusyellowl}  & Sensitivity (95\% CI)  & 0.778 (0.750–0.806) & 1.000 (1.000–1.000) \\
 \rowcolor{cusyellowl}  & Specificity (95\% CI)  & 0.895 (0.874–0.916) & 0.918 (0.893–0.943) \\

\hline 
\end{tabular} 
\label{ST_wsi_3}
\end{table}

\clearpage
\begin{table}[h] 
\renewcommand{\arraystretch}{2}
\renewcommand{\tablename}{Extended Data Table.}
\centering 
\caption{\textbf{External testing results of Smart-CCS in the retrospective study.} Base refers to the typical two-step CCS model, w/ P is adding pretraining, w/ P\&A denotes our proposed Smart-CCS with pretraining and adaptation.}
\begin{tabular}{cc|ccc} 
\hline
 \rowcolor{cusyellow} \multicolumn{2}{c|}{\multirow{1}{*}{\textbf{Metrics}}} & \textbf{Base} &\textbf{w/ P }&\textbf{w/ P\&A (ours)}\\ 
\hline
\multirow{1}{*}{RC8}&Accuracy (95\% CI) & 0.680 (0.650–0.711)&0.802 (0.776–0.828)&0.837 (0.820–0.854)\\
& AUC (95\% CI) &0.812 (0.787–0.837) & 0.851 (0.828–0.874)& 0.923 (0.911–0.935)\\
& F1 (95\% CI) &0.750 (0.722–0.778)&0.840 (0.816–0.864) &0.867 (0.851–0.882)\\
& Sensitivity (95\% CI) &0.762 (0.735–0.790)& 0.775 (0.748–0.802) & 0.891 (0.877–0.905)\\
& Specificity (95\% CI) &0.673 (0.642–0.703)& 0.805 (0.779–0.830)&0.832 (0.815–0.849)\\

 \rowcolor{cusyellowl}  \multirow{1}{*}{RC9}&Accuracy (95\% CI) & 0.812 (0.785–0.838)&0.897 (0.876–0.918)&0.938 (0.927–0.949)\\
 \rowcolor{cusyellowl} & AUC (95\% CI) &0.897 (0.877–0.918) &0.930 (0.913–0.948)&0.963 (0.954–0.972)\\
 \rowcolor{cusyellowl} & F1 (95\% CI) & 0.823 (0.797–0.849) &0.898 (0.878–0.919)&0.937 (0.926–0.949) \\
 \rowcolor{cusyellowl} & Sensitivity (95\% CI) & 0.821 (0.795–0.847) &0.816 (0.789–0.842) &0.833 (0.816–0.851)\\
 \rowcolor{cusyellowl} & Specificity (95\% CI) & 0.809 (0.783–0.836)&0.921 (0.903–0.939) &0.967 (0.958–0.975)\\

\multirow{1}{*}{RC10}&Accuracy (95\% CI) &0.741 (0.712–0.770) &0.871 (0.849–0.893) &0.906 (0.892–0.919)\\
& AUC (95\% CI) &0.830 (0.805–0.855)&0.911 (0.892–0.930) &0.953 (0.943–0.963)\\
& F1 (95\% CI) &0.771 (0.743–0.799)& 0.879 (0.857–0.901)& 0.911 (0.898–0.924)\\
& Sensitivity (95\% CI) &0.781 (0.753–0.808)& 0.808 (0.782–0.834) &0.892 (0.878–0.907) \\
& Specificity (95\% CI) &0.733 (0.704–0.762)&0.884 (0.863–0.905)& 0.909 (0.895–0.922) \\

 \rowcolor{cusyellowl}  \multirow{1}{*}{RC11 }&Accuracy (95\% CI) & 0.815 (0.788–0.841) &0.886 (0.864–0.907)& 0.921 (0.908–0.934)\\
 \rowcolor{cusyellowl} & AUC (95\% CI) & 0.875 (0.853–0.898)&0.935 (0.919–0.952)&0.962 (0.953–0.971)\\
 \rowcolor{cusyellowl} & F1 (95\% CI) &0.815 (0.789–0.841) &0.887 (0.865–0.908) &0.921 (0.908–0.933)\\
 \rowcolor{cusyellowl} & Sensitivity (95\% CI) & 0.761 (0.732–0.790)   &0.895 (0.874–0.916) &0.871 (0.855–0.887)\\
 \rowcolor{cusyellowl} & Specificity (95\% CI) & 0.846 (0.822–0.871)  &0.880 (0.859–0.902)&0.951 (0.941–0.961)\\

\multirow{1}{*}{RC12}&Accuracy (95\% CI) & 0.804 (0.771–0.836)  &0.828 (0.797–0.859)&0.871 (0.851–0.891)\\
& AUC (95\% CI) &0.871 (0.843–0.898)& 0.916 (0.894–0.939) &0.941 (0.927–0.955) \\
& F1 (95\% CI) &0.804 (0.772–0.836)& 0.826 (0.796–0.857) & 0.871 (0.851–0.891)\\
& Sensitivity (95\% CI) &0.786 (0.752–0.819)&0.899 (0.875–0.924)& 0.884 (0.865–0.903)\\
& Specificity (95\% CI) &0.824 (0.793–0.855) & 0.747 (0.712–0.783)  &0.857 (0.836–0.878)\\

 \rowcolor{cusyellowl}  \multirow{1}{*}{RC13}&Accuracy (95\% CI) & 0.644 (0.522–0.766)&0.793 (0.761–0.825)& 0.828 (0.804–0.852)\\
 \rowcolor{cusyellowl} & AUC (95\% CI) & 0.635 (0.513–0.758)  &0.815 (0.784–0.846) &0.880 (0.859–0.901)\\
 \rowcolor{cusyellowl} & F1 (95\% CI) & 0.644 (0.522–0.766) &0.814 (0.783–0.845)&0.846 (0.823–0.869)\\
 \rowcolor{cusyellowl} & Sensitivity (95\% CI) & 0.659 (0.621–0.697)&0.950 (0.939–0.960)&0.771 (0.744–0.798)\\
 \rowcolor{cusyellowl} & Specificity (95\% CI) & 0.594 (0.468–0.719) &0.816 (0.785–0.847) &0.838 (0.814–0.861)\\

\hline 
\end{tabular} 
\label{ST_ext}
\end{table}

\clearpage
\begin{table}[h] 
\renewcommand{\arraystretch}{2}
\renewcommand{\tablename}{Extended Data Table.}
\centering 
\caption{\textbf{Performance comparisons of different classification methods in WSI classification task (AUC (95\% CI)).} The best results are in bold.}
\begin{tabular}{c|cc} 
\hline
 \rowcolor{cusyellow} \textbf{Methods} & \textbf{Internal Test}  & \textbf{External Test} \\ 
\hline
MeanMIL & 0.965 (0.961–0.969)&0.950 (0.945–0.954) \\
 \rowcolor{cusyellowl} MaxMIL & 0.957 (0.953–0.961) &0.933 (0.928–0.938)\\
ABMIL \cite{ilse2018attention} & 0.965 (0.961–0.969) &0.947 (0.942–0.952) \\
 \rowcolor{cusyellowl} DSMIL \cite{li2021dual} & 0.959 (0.955–0.964) & 0.938 (0.933–0.943)  \\
CLAM-SB \cite{shao2021transmil} & 0.964 (0.960–0.968) & 0.919 (0.914–0.925) \\
 \rowcolor{cusyellowl}  TransMIL \cite{lu2021data} & 0.966 (0.962–0.970)  & 0.951 (0.946–0.955)\\
S4MIL \cite{fillioux2023structured} &\textbf{0.969 (0.966–0.973)} & \textbf{0.953 (0.948–0.957)}\\
\hline 
\end{tabular} 
\label{ST_clas}
\end{table}

\clearpage
\begin{table}[h] 
\renewcommand{\arraystretch}{2}
\renewcommand{\tablename}{Extended Data Table.}
\centering 
\caption{\textbf{Performance of cervical cytology cancer screening in the prospective study across three centers (PC1-PC3).}}
\begin{tabular}{c|ccc} 
\hline
 \rowcolor{cusyellow} \textbf{Metrics} & \textbf{PC1 ($N$=998)} & \textbf{PC2 ($N$=1,311)} & \textbf{PC3 ($N$=1,044)}\\ 
\hline
Accuracy (95\% CI) & 0.877 (0.856–0.897)&0.862 (0.843–0.881) &0.950 (0.937–0.963)\\
 \rowcolor{cusyellowl} AUC (95\% CI) &0.947 (0.933–0.961)   & 0.924 (0.910–0.938)& 0.986 (0.979–0.993) \\
F1 Score (95\% CI) & 0.877 (0.857–0.898) & 0.867 (0.848–0.885)   &0.951 (0.938–0.964)\\
 \rowcolor{cusyellowl} Sensitivity (95\% CI) &0.893 (0.874–0.912) & 0.881 (0.864–0.899) & 0.946 (0.932–0.960)\\
Specificity (95\% CI) &0.864 (0.843–0.885)& 0.855 (0.836–0.874)  &0.952 (0.939–0.965)\\

\hline 
\end{tabular} 
\label{ST_pros}
\end{table}

\newpage
\begin{table}[h] 
\renewcommand{\arraystretch}{1.5}
\renewcommand{\tablename}{Extended Data Table.}
\centering 
\caption{\textbf{Statistics of the CCS-127K dataset with the number of WSIs and corresponding patches at each center.}}
\begin{tabular}{c|c|c|c|c|c} 
\hline
 \rowcolor{cusyellow} \textbf{Center} & \textbf{WSI} & \textbf{Patch} & \textbf{Center} & \textbf{WSI} & \textbf{Patch} \\
\hline
 RC1 & 12,598 & 26,980,754   & RC26 & 60 & 63,215\\
 \rowcolor{cusyellowl}  RC2 & 14,529 & 17,400,701&  RC27 & 239 & 102,381\\
 RC3 & 9,039  & 27,416,532 & RC28 & 33& 22,435\\
 \rowcolor{cusyellowl}  RC4 &8,432 & 14,811,441 &  RC29 & 20 & 23,921 \\
 RC5 & 17,623 & 33,845,451& RC30 & 13 & 63,745\\
 \rowcolor{cusyellowl}  RC6 & 5,773 & 15,692,295 &  RC31 & 10& 5,527\\
 RC7 & 8,139 & 31,573,365 & RC32 & 5,516 & 2,290,698\\
 \rowcolor{cusyellowl}  RC8 & 1,873 & 2,002,922&  RC33 & 160 &354,062\\
 RC9 & 2,013 & 6,085,661 & RC34 & 1,319 & 1,732,829\\
 \rowcolor{cusyellowl}  RC10 &  1,797 & 1,776,018 &  RC35 & 543 & 1,921,444 \\
 RC11 & 3,400 & 8,420,141 & RC36 & 491 & 2,970,771\\
 \rowcolor{cusyellowl}  RC12 & 1,097 & 1,867,367&  RC37 & 6& 10,337\\
 RC13 &  1,876 & 371,098 & RC38 & 39 & 42,763 \\
 \rowcolor{cusyellowl}  RC14 & 12,823 & 12,832,172 &  RC39 & 13& 14,433\\
 RC15 & 657 & 979,460 & RC40 & 127 & 181,802 \\
 \rowcolor{cusyellowl}  RC16 &588 & 68,299 &  RC41 & 190 & 63,359\\
 RC17 & 707 & 231,009 & RC42 & 7,002 & 9,228,021\\
 \rowcolor{cusyellowl}  RC18 & 491 & 451,371 &  RC43 & 330 & 54,372\\
 RC19 &494 & 300,671 &  RC44 & 451 & 428,447 \\
 \rowcolor{cusyellowl}  RC20 & 933 & 1,749,139 &  RC45 & 14 & 45,227 \\
 RC21 & 234 & 85,197 & PC1 & 998 & 421,901\\
 \rowcolor{cusyellowl}  RC22 & 2,045 & 372,867 &  PC2 & 1,311 & 788,410\\
 RC23 & 192 & 223,526 & PC3 &1,044 & 602,583\\
 \rowcolor{cusyellowl}  RC24 & 108 & 51,613&  Total & 127,471& 227,266,352\\
RC25 & 81 & 244,599& & & \\
\hline
\end{tabular} 
\label{ST_patches}
\end{table}

\newpage
\begin{table}[h] 
\renewcommand{\arraystretch}{1.5}
\renewcommand{\tablename}{Extended Data Table.}
\centering 
\caption{\textbf{Hyperparameters used for self-supervised pretraining.}}

\begin{tabular}{c|c|p{2cm}c} 
\hline 
 \rowcolor{cusyellow} & \textbf{Hyperparameters} & \textbf{Value} \\
\hline 
\multirow{7}{*}{Model} & Layer number  & 24 \\
& Feature dimension & 1,024 \\
& Patch size & 14 \\
& Heads number & 16 \\
& FFN layer & MLP \\
& Drop path ratio & 0.4 \\
& Layer scale  & 1.00e-05 \\
\hline 
 \rowcolor{cusyellowl} Loss weight & DINO & 1 \\
 \rowcolor{cusyellowl}& iBOT & 1 \\
\hline 
\multirow{12}{*}{Optimization} & Teacher momentum & 0.994\\
& Total batch size & 1,024 \\
& Base learning rate & 1.00e-04 \\
& Minimum learning rate & 1.00e-06 \\
& Global crops scale & 0.32, 1.0 \\
& Global crops size & 224 \\
& Local crops scale & 0.02, 0.32 \\
& Local crops size & 98 \\
& Local crops number & 8 \\
& Gradient clip & 3\\
& Warmup iterations & 50,000 \\
& Total iterations & 500,000\\ 
\hline 
\end{tabular} 
\label{ST_pretrain_para}
\end{table}

\clearpage
\begin{table}[h] 
\renewcommand{\arraystretch}{1.5}
\renewcommand{\tablename}{Extended Data Table.}
\centering 
\caption{\textbf{Hyperparameters used for cell detector and WSI classifier in CCS model.}}
\begin{tabular}{c|c} 
\hline
 \rowcolor{cusyellow} \multicolumn{2}{c}{\multirow{1}{*}{\textbf{Abnormal Cell Detector}}} \\ 
\hline
backbone & ResNet50\\ 
 \rowcolor{cusyellowl} num\_encoder\_layers & 6\\ 
num\_decoder\_layers & 6\\ 
 \rowcolor{cusyellowl} num\_queries & 300\\ 
num\_classes & 7\\ 
 \rowcolor{cusyellowl} dropout & 0.1\\ 
epoch & 100\\ 

\hline
 \rowcolor{cusyellow} \multicolumn{2}{c}{\multirow{1}{*}{\textbf{WSI Classifier}}} \\ 
\hline
classifier & ABMIL, MeanMIL, MaxMIL, CLAM, DSMIL, TransMIL, S4MIL\\ 
 \rowcolor{cusyellowl} top-k & 50\\ 
epoch & 50\\ 
 \rowcolor{cusyellowl} num\_classes & 7\\ 
in\_dim & 1,024\\ 
\hline 
\end{tabular} 
\label{ST_det_para}
\end{table}